\documentclass[5p,twocolumn,10pt]{elsarticle} 

\usepackage{amsmath}
\usepackage{hyperref}
\usepackage[toc,page]{appendix}

\usepackage{lineno}
\usepackage{hyperref}
\usepackage{enumitem}
\modulolinenumbers[5]

\usepackage{subcaption}
\usepackage{mathtools}
\usepackage{amsmath}
\usepackage{amssymb}
\usepackage{stmaryrd}
\usepackage{amsthm}
\usepackage{amsfonts}
\usepackage{dsfont}
\usepackage{graphicx}
\usepackage{upgreek}
\usepackage{bm}
\usepackage{algpseudocode}
\usepackage{algorithm}
\usepackage{caption}
\usepackage{subcaption}
\usepackage{color}
\usepackage{float}
\usepackage{scrextend}
\usepackage[cal=boondoxo]{mathalfa}
\usepackage{nicefrac}
\usepackage{algorithm,algpseudocode} 
\usepackage{todonotes}

\biboptions{numbers,sort&compress}

\DeclareFontFamily{OT1}{pzc}{}
\DeclareFontShape{OT1}{pzc}{m}{it}{<-> s * [1.200] pzcmi7t}{}
\DeclareMathAlphabet{\mathpzc}{OT1}{pzc}{m}{it}

\newcommand{\id}{\mathsf{id}}

\newcommand{\bx}{{\mathbf{x}}}

\newcommand{\bk}{{\mathbf{k}}}

\newcommand{\bt}{{\mathbf{t}}}

\newcommand{\be}{{\mathbf{e}}}

\newcommand{\sign}{\mathsf{sign}}

\newcommand{\indic}{{\mathbf{1}}}

\newcommand{\eqae}{\overset{\mathsf{ae}}{=}}

\newcommand{\R}{{\mathds{R}}}

\newcommand{\SO}[1]{{\mathrm{SO}(#1)}}

\newcommand{\conf}{{\mathsf{C}}}

\newcommand{\OC}{\mathsf{OC}}
\newcommand{\UC}{\mathsf{UC}}

\renewcommand{\i}{\mathsf{i}}
\renewcommand{\o}{\mathsf{o}}

\theoremstyle{definition}

\newcommand{\eq}[1]{(\ref{#1})} 
\newcommand{\com}[1]{} 

\journal{Computer-Aided Design}

\begin{document}
	
\baselineskip11pt
\begin{frontmatter}

\title{Hybrid Manufacturing Process Planning for Arbitrary Part and Tool Shapes}

\author{George Harabin and Morad Behandish}

\address{\rm
	Palo Alto Research Center (PARC),
	3333 Coyote Hill Road, Palo Alto, California 94304
	\vspace{-15.0pt}
}

\begin{abstract}
	Hybrid manufacturing (HM) technologies combine additive and subtractive manufacturing (AM/SM) capabilities in multi-modal process plans that leverage the strengths of each. Despite the growing interest in HM technologies, software tools for process planning have not caught up with advances in hardware and typically impose restrictions that limit the design and manufacturing engineers' ability to systematically explore the full design and process planning spaces. We present a general framework for identifying AM/SM actions that make up an HM process plan based on accessibility and support requirements, using morphological operations that allow for arbitrary part and tool geometries to be considered. To take advantage of multi-modality, we define the actions to allow for temporary excessive material deposition or removal, with an understanding that subsequent actions can correct for them, unlike the case in unimodal (AM-only or SM-only) process plans that are monotonic. We use this framework to generate a combinatorial space of valid, potentially non-monotonic, process plans for a given part of arbitrary shape, a collection of AM/SM tools of arbitrary shapes, and a set of relative rotations (fixed for each action) between them, representing build/fixturing directions on $3-$axis machines. Finally, we define a simple objective function quantifying the cost of materials and operating time in terms of deposition/removal volumes and use a search algorithm to explore the exponentially large space of valid process plans to find ``cost-optimal'' solutions. We demonstrate the effectiveness of our method on 3D examples.
\end{abstract}


\end{frontmatter}

\section{Introduction} \label{sec_intro}

In recent years, additive-manufacturing (AM) technologies have enabled the manufacture of parts with unprecedented levels of geometric/material complexity and mass customization. More recently, post-pandemic supply chain challenges has led to a surge in investment in AM as a means to produce on-demand spare parts at the point of need. However, compared to subtractive manufactruing (SM) processes (such as CNC machining), most AM processes suffer from diminished surface quality and reduced dimensional accuracy to varying degrees \cite{Rifat2022effect, Gong2020machining}, in a complex tradeoff with workspace size, build time, energy, and cost, among other factors. On the other hand, SM processes cannot provide the same levels of geometric complexity due to tool accessibility constraints and result in more material waste compared to additive processes \cite{Basinger2018development}. Hybrid manufacturing (HM)  offers tremendous potential to combine the relative strengths of both AM and SM approaches in a multi-modal sequence of manufacturing actions that leverage the best of both worlds. Examples of HM use-cases consist of directed energy deposition (DED) and powder bed fusion (PBF), combined with CNC machining to perform coating and repair \cite{Cortina2018latest}. See \cite{Sealy2020hybrid, Popov2020hybrid, Pragana2021hybrid} for recent reviews of HM technologies.

\begin{figure} [ht!]
	\centering
	\includegraphics[width=0.65\linewidth]{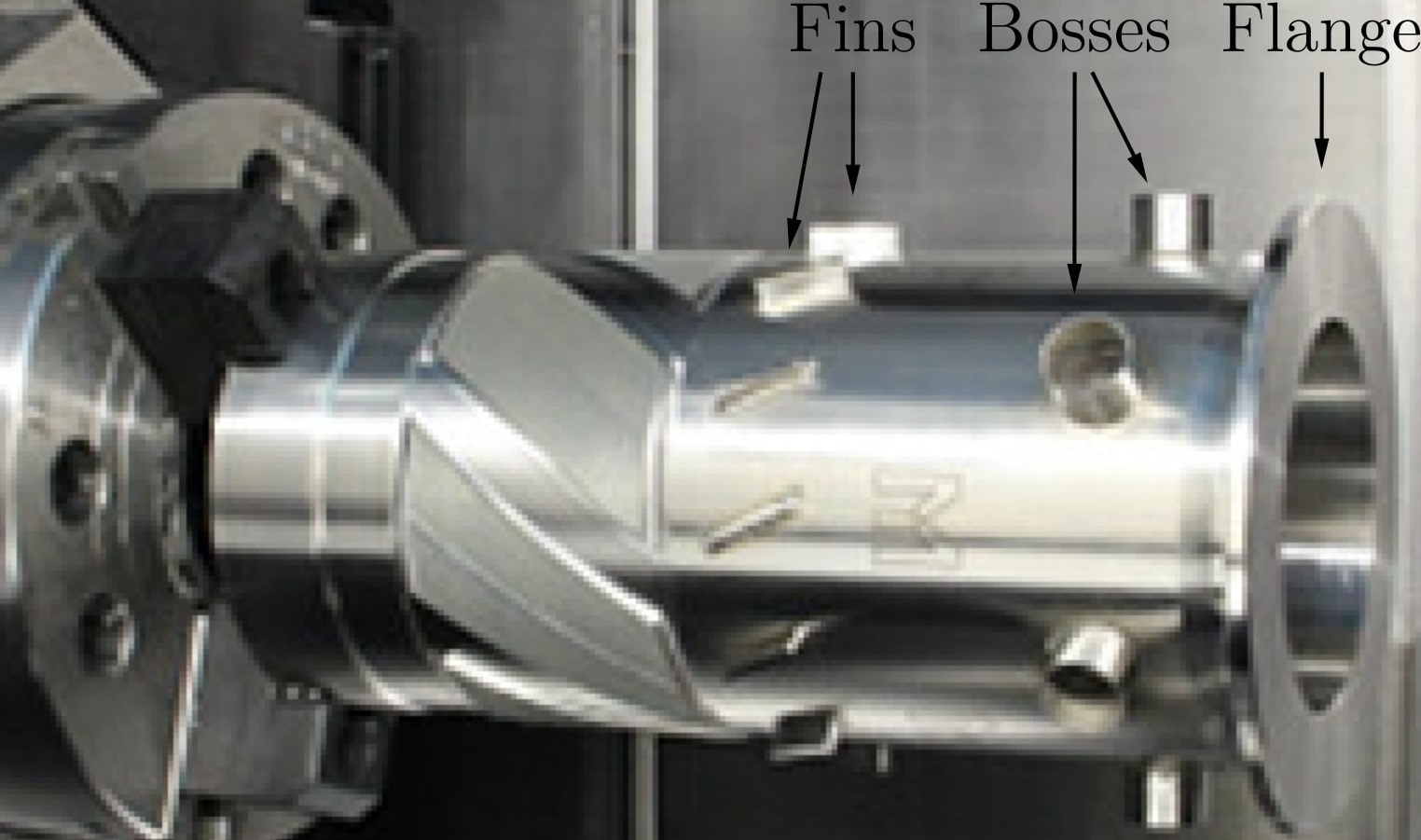}
	\caption{Part produced with 3 distinct features added via AM-then-SM pairs of actions on a \textsf{Mazak INTEGREX i-400 AM} \cite{Behandish2018automated,Yamazaki2016development}.}
	\label{fig_Mazac}
\end{figure}

While HM has overcome some of the limitations of uni-modal (AM-only or SM-only) processes, its industrial use-cases have remain limited to modifying parts with easily separable features for AM/SM actions based on shape or materials (Fig. \ref{fig_Mazac}). The true potential of HM in producing parts can be unlocked by HM-specialized computer-aided process planning (CAPP) software tools. Current approaches to HM process planning, however, rely on assumptions about part geometry or use feature based decompositions that limit their extensibility to complex part and tool geometries  \cite{Liu2019sequence, Chen2019manufacturability, Zheng2020cost, Basinger2018development}. 

As discussed in our earlier work \cite{Behandish2018automated}, HM process planning faces several theoretical and algorithmic challenges that are not present in uni-modal (AM-only or SM-only) process planning, such as a) {\it non-monotonicity}, e.g., material may be removed only to be added back in at a later point in the plan, and vice versa; and b) {\it non-permutativity}, i.e., changing the order of a pair of AM-SM actions with overlapping regions of influence (ROI) in the 3D workspace affects the end result, unlike the case with AM-AM or SM-SM pairs. As a result of such impediments to HM process planning, currently practical HM use-cases are primarily restricted to short (e.g., two-step) sequences. For example, a common scenario is an AM-then-SM sequence consisting of an initial AM operation to deposit material into a ``near-net'' shape, followed by an SM operation intended to remove excess material (e.g., sacrificial support structures) and finish functional surfaces down to desired error tolerance and quality \cite{Karunakaran2010low, Zhang2020development}. Another common scenario is an SM-then-AM sequence consisting of an initial SM operation to carve out cracked regions of a high-value part (e.g., a turbine blade), followed by an AM operation intended to fill in the removed region with new material \cite{Ren2007part, Wilson2014remanufacturing}.

This paper presents a novel approach to HM process planning that can unlock the full potential of HM in producing complex geometries from scratch. We construct process plans by iteratively adding AM/SM actions (when needed) to get an initial state---e.g., a raw/bar stock, if the first action is SM, or an empty build plate, if the first action is AM---to a desired target state, while minimizing cost. Our approach enables identifying cost-effective HM process plans to build complex geometries by strategic interleaving of AM and SM actions. Such computational tools can accelerate the industrial adoption of AM in concert with SM, leveraging their complementary advantages.

\subsection{Related Work} \label{sec_lit}

Commercial-strength software tools for HM processes (e.g., \textsf{CGTech VERICUT} or \textsf{Siemens NX}) are mostly focused on feature-based path planning and process simulation for specific processes and motion controllers, extending traditional computer-aided manufacturing (CAM) for CNC tooling. Higher-level CAPP activities such as identifying how to break a given part's interior/exterior to AM/SM volumes, mapping them to proper AM/SM capabilities, and ordering the AM/SM actions for cost optimization, are still in academic stage. Previous studies have aimed to address the limitations of CAPP tools through various approaches such as feature-based decomposition of the part or by making simplifying assumptions about part and/or tool shapes. 

Liu et al. \cite{Liu2019sequence} presented a decomposition-based approach to sequence planning for multi-axis HM of geometrically complex parts, incorporating tool accessibility through a projection-based approach. Chen et al. \cite{Chen2019manufacturability} examined the problem of HM process planning for $2.5-$axis AM, using a sequential search algorithm to discover a minimal length alternating HM process plan (switching AM/SM modalities at each step) while accounting for tool accessibility constraints. The part geometries for both of these approaches, however, were limited to columnar shapes. 

Zheng et. al \cite{Zheng2020cost} proposed an approach to cost-driven process planning formulated as mixed-integer programming, relying on feature-based decomposition of the part in order to build the cost model. Basinger et al. \cite{Basinger2018development} developed the feature-based hybrid manufacturing process planning system (FAH-PS) that utilized feature-specific geometric, tolerance, and material data in order to automatically generate HM process plans that minimize tool and part orientation changes. However, making simplifying assumptions about geometry limits the HM search space, while feature-based decomposition relies on engineering intent and limits extensibility to parts and tools of complex shapes \cite{Behandish2018automated}.

Several approaches have combined the development of novel HM technologies with custom-designed process planning software targeted towards the process: Manogharan et al. \cite{Manogharan2015aims} developed an integrated system (called \textsf{AIMS}) to integrating direct metal AM techniques (e.g., EBM or DMLS) with CNC machining (for creating critical surfaces and dimensions) in order to produce an HM system which is capable of manufacturing metal parts with complex shapes while maintaining accuracy and good surface finish. The \textsf{AIMS} system relied on a collection of tools used in pure AM/SM process planning, including visibility analysis, fixture design and location, maximum deviation analysis, surface overgrowth and tool-path analysis.

Karunakaran et al. \cite{Karunakaran2010low} developed the two-step layered \textsf{ArcHLM} method, which consisted of weld-deposition of a layer followed by a finish machining step, along with software to generate weld and face milling paths. Process planning for \textsf{ArcHLM}, however, is limited to a repeated two-step sequence of AM-then-SM steps which bring the part to a near-net shape, followed by face-milling the deposited layer to the required thickness in order to provide good surface quality to facilitate the deposition of the next layer.

Newman et. al \cite{Zhu2012novel} developed the \textsf{iAtractive} HM process, which is targeted at reuse and remanufacture of existing and legacy parts. The \textsf{iAtractive} process consists of fused-filament AM combined with CNC machining and an inspection step, along with a ten-step reactionary process planning approach (R$\text{P}^2$A). This process utilizes a combination of feature extraction, interpretation, and modification along with part decomposition in order to analyze machinability and printability of a part and determine build direction, operation sequencing, and tool-path generation for the optimal process plan \cite{Zhu2013development, Newman2015process}.

Due to their reliance on feature extraction or limiting assumptions about the geometry and the process, many of the previous approaches are restricted to relatively simple part and/or tool shapes and do not allow for systematic exploration of the entire process planning space \cite{Mirzendehdel2019exploring}. 

Behandish et al. \cite{Behandish2018automated}, inspired by earlier work in SM-only process planning by Nelaturi et al. \cite{Nelaturi2015automatic}, aimed to lift these restrictions by presenting a generic definition of an HM process plan as an arbitrary sequence of AM/SM actions abstracted using shape-agnostic logical and set-theoretic constructs. Each action was defined by its ROI in the 3D workspace, computed using concepts from mathematical morphology \cite{Serra1983image,Roerdink2000group} and configurations space ($\conf-$space) modeling \cite{Lozano-Perez1983spatial}. The key advantage of this approach is its ability to perform purely logical process planning in the finite Boolean algebra (FBA) of precomputed ROIs (or their point membership tests) and their canonical intersections \cite{Shapiro1997maintenance}, decoupled from spatial reasoning that generate the ROI shapes. The main limitations of the approach were its neglection of the evolution of $\conf-$space obstacle, which may affect tool accessibility, and the use of conservative policies for developing ROI candidates, which miss the opportunity to realize the full potential of HM, as elaborated in Section \ref{sec_back}. In this paper, we overcome these limitations by embedding spatial reasoning back into the process planning and using conservative (under-fill/over-cut) and aggressive (over-fill/under-cut) policies, detailed in Section \ref{sec_revisit}.

\subsection{Contributions \& Outline}

This paper presents a general methodology to automatically generate HM process plans for a given part of arbitrary target shape and given set of AM/SM capabilities with $3-$axis motion and arbitrary tool shapes. Our key contributions are summarized as follows:
\begin{enumerate}
	\item To define a discrete search space, we conceptualize AM/SM actions as transformations of the workpiece according to conservative and aggressive policies.
	\item We use morphological operations and $\conf-$space reasoning to compute these transformations, subject to accessibility (collision avoidance, for both AM and SM) and support requirements (overhang avoidance, for AM).
	\item We use an iterative deepening A* (IDA*) search algorithm on this search space to identify cost-effective process plans with respect to a cost function that quantifies ``wasted'' (i.e., deposited and subsequently removed) material throughout the process.
\end{enumerate}

The {\it conservative} policy is defined by ``maximal'' under-filling (AM) and over-cutting (SM), i.e., getting as close as possible to the target shape in a single greedy action, presented in \cite{Behandish2018automated} inspired by \cite{Nelaturi2015automatic}. The {\it aggressive} policy, on the other hand, aims to exploit the unique advantage of multi-modal HM over uni-modal (AM-only or SM-only) process plans; namely, over-filling (AM) and under-cutting (SM) with ``minimal'' sacrificial deviation from the target shape to enable the action according to accessibility and support requirements. The search space is thus constructed as a tree, whose branching factor is determined by the number of AM/SM capabilities, with two policy options per each.

To define workpiece transformations, we define AM/SM actions as Boolean union/difference operations between the intermediate workpiece shape and the actions' material deposition/removal ROIs, respectively:
\begin{itemize}
	\item For SM, in the absence of an ordering on the motion sets (e.g., a tool-path), we use rigorous set-theoretic equations to {\it implicitly} define accessible regions in a self-referencing manner. The removable region is defined as the fixed point of a recursive definition, solved iteratively until convergence is achieved. 
	\item For AM, while accessibility is straightforward to compute (without the need for iteration), the motion is further constrained by overhang avoidance, defined by downward projection (i.e., ray tracing along gravity)
\end{itemize}
The set-theoretic formulations of accessibility (i.e., collision avoidance) and support (i.e., overhang avoidance) constraints are given in terms of Minkowski operations and computed using fast Fourier transform (FFT) accelerated convolutions on voxel representations.

Section \ref{sec_overview} presents an overview of the approach, including the groundwork from \cite{Behandish2018automated}, its limitations, and how we address them in this paper differently. Section \ref{sec_geom} presents the basic geometric operations for accessibility, collateral, and support analysis. Section \ref{sec_actions} presents the definition and computation of AM/SM actions, using the said geometric operations. In both of these sections, we present general set-theoretic formulations that are independent of the choice of representation scheme (e.g., B-reps or mesh). For computational purposes, we also present efficient algorithms for voxel-based representations. Section \ref{sec_search} presents the search space, defined in terms of actions, the cost function, and the IDA* search algorithm for cost-effective HM process planning. Section \ref{sec_results} presents a few examples in 3D that demonstrate the effectiveness of our approach. We conclude the paper with final remarks in Section \ref{sec_conclusion}.

\section{Overview of the Approach} \label{sec_overview}

In this section, we review our earlier work in greater detail along with its limitations (Section \ref{sec_back}) and illustrate how the new approach in this paper can overcome those limitations (Section \ref{sec_revisit}).

\subsection{Background} \label{sec_back}

In \cite{Behandish2018automated}, we presented a set-theoretic formalism of HM process plans with arbitrary part and tool shapes. The evolving state of a workpiece was modeled by a finite sequence of 3D pointsets modified by AM/SM `actions'. Each action was modeled by a set union/difference operation with a corresponding AM/SM `primitive', carried out by a single AM/SM `capability', respectively. An AM/SM capability was defined formally by the available machine degrees-of-freedom (DOF) and minimum manufacturable neighborhood (MMN). The corresponding AM/SM primitive could be any region of space that is producible by {\it sweeping} an MMN (e.g., solidified liquid droplets or melt-pools for certain AM and milling/turning tool inserts for certain SM) along an available motion, restricted by DOF and collision avoidance (i.e., {\it accessibility}) constraints. The AM/SM primitives could be viewed as overlapping volumes that eventually make up the interior/exterior of the target shape, respectively, as the workpiece evolves, i.e., their proper Boolean combination ultimately makes the final shape. 

Our formulation of HM actions and process plans in \cite{Behandish2018automated} was inspired by (and extended) the ideas in \cite{Nelaturi2015automatic}, which applied to uni-modal machining process planning, and viewed an SM process plan as a sequence of SM actions that took out maximal removable volumes with given tools at different orientations. These volumes were computed for arbitrary part and tool shapes using morphological operations \cite{Serra1983image}, which can be extended from Euclidean $3-$space (e.g., for $3-$axis milling) to higher-dimensional configuration spaces \cite{Lozano-Perez1983spatial} (e.g., for high-axis turn-mill) using group morphology \cite{Roerdink2000group,Lysenko2010group}. Due to the uni-modal nature of SM-only process plans, the ``maximality'' is desirable as a {\it greedy} policy to take out as much material as possible in a given fixturing setup for a given DOF and tool geometry. These process plans are monotonic, meaning that a removed region by one action cannot be brought back by a subsequent action, which leads to a precise definition of maximality that must constrain all SM actions. Moreover, these process plans are permutative, meaning that the final shape can be computed by unifying all removable volumes, regardless of the order in which they apply, and subtracting this union from the raw stock. This realization enabled a neat separation of manufacturability analysis (i.e., checking if the target shape is achievable by order-agnostic Boolean operations) and process planning (i.e., picking an order to optimize cost).

In \cite{Behandish2018automated}, we defined maximal depositable/removable region (MDR/MRR) for generalized HM capabilities defined by arbitrary motion DOF and MMN shapes, implemented for $3-$axis translational DOF. However, we encountered a number of challenges to perform multi-modal manufacturability analysis and process planning. 

First, using MDR/MRR as AM/SM primitives for HM, respectively, does not leverage the added freedom in non-monotonic process plans, i.e., there is no reason to constrain the actions by maximality. In fact, the true power of HM can be realized only if the actions are allowed to over-fill/under-cut with respect to the target shape, with the ability to compensate by subsequent actions. Hence, we need more sophisticated policies than the greedy approach, allowing for such actions. In this paper, we define two policies to choose from for every AM/SM action; namely, a {\it conservative} policy that uses ``maximal'' under-fill/over-cut regions with respect to the target shape, as before, and a new {\it aggressive} policy that allows for over-fill/under-cut with ``minimal'' sacrificial deviation from the target shape, assuming it can be compensated for by future actions of opposite kind. We consider accessibility requirements for SM, by either avoiding collisions with the target shape or incurring collateral damage to make the inaccessible regions accessible. We also consider accessibility requirements for AM, by avoiding collisions with the current workpiece shape, as well as support requirements for AM, by either avoiding overhangs or adding sacrificial support material for AM.

Second, the loss of permutativity in multi-modal process plans makes it difficult to decouple manufacturability analysis from process planning for HM. Unlike uni-modal process plans, defined by order-agnostic Boolean formulae, multi-modal process plans are order-sensitive. In other words, the final shape cannot be determined from a given set of mixed AM/SM primitives without specifying the order in which they are applied. To mitigate this challenge in \cite{Behandish2018automated}, we employed a canonical decomposition approach \cite{Shapiro1997maintenance} to generate ``atomic'' regions of space from intersections of AM/SM primitives and their complements. HM manufacturability analysis was thus reduced to testing whether any subcollection of atoms, partitioning the 3D space, makes up the final shape (necessary condition) and if the disjunctive normal form (DNF) implied by this subcollection is equivalent to some Boolean formula in terms of AM/SM primitives that represents valid process plans (sufficient condition). A key advantage of this formulation was that all logical reasoning occurred at a granularity determined by atoms rather than much smaller geometric representation units (e.g., voxels). It also enabled faster process planning via purely combinatorial or logical manipulation of the finite Boolean formulae (i.e., no geometric computations in the loop).

However, a major limitation of the above decoupling was that it could not account for the evolution of $\conf-$space obstacles. The $\conf-$space analysis used to precompute collision-free AM/SM primitives that satisfy accessibility was based on the target shape. Hence, the AM/SM actions defined as Boolean operations with such primitives may not be applicable to the intermediate shapes of the workpiece, as the $\conf-$space obstacle evolves as well. For example, for an SM action to be applicable, there was an implicit assumption in \cite{Nelaturi2015automatic,Behandish2018automated} that the regions in an intermediate workpiece that will not end up in the desired target shape---potentially introducing additional accessibility constraints in the intermediate actions---can be taken out by the same SM action whose accessibility is being evaluated. 

In this paper, we show that such an assumption is not always valid and a proper analysis requires an iterative process to solve a set-theoretic self-referencing definition. Moreover, we embed such geometric computations within the process planning to ensure accurate results with respect to evolving accessibility conditions. Although this embedding may result in a loss of computational efficiency (due to geometric computations in the loop) compared to the purely logical reasoning proposed in \cite{Behandish2018automated}, the process planning can be sped up using iterative deepening techniques, akin to what was originally used in \cite{Nelaturi2015automatic}.

\subsection{Manufacturing Actions Revisited} \label{sec_revisit}

Our approach to generating valid HM process plans closely follows the definition of manufacturing primitives given in \cite{Behandish2018automated} as ``the best shot one can take using a single (either AM or SM) capability, to transition from a given state of the workpiece to a state that is closer to the target state, while respecting the manufacturing constraints of that capability.'' With this principal in mind, we seek to generate  process plans through a concatenation of valid AM/SM actions, which are either conservative or aggressive in their addition/removal of material, respectively. 

For SM, `valid' conservative and aggressive actions are defined, respectively, as follows:
\begin{itemize}
	\item An over-cut (OC) action is a one-shot material removal operation that removes as much excess material as possible from a given workpiece state without removing any region that is inside the target state.
	\item An under-cut (UC) action is a one-shot material removal operation that removes the entire excess material from the given workpiece state, regardless of whether it requires removal of additional regions that are inside the target state.
\end{itemize}
The ``excess'' material in both definitions refers to material that is inside the current state but outside the target state. The removal of excess material and/or additional regions are constrained by accessibility and MMN shape. More precisely, for a given target shape $P \subset \R^3$, current workpiece shape (input state) $P_\i \subset \R^3$, and tool shape $T\subset \R^3$, the OC action removes as much of the excess material (i.e., maximal subset of $P_\i - P$) that is accessible by $T$, without having to remove anything in $P \cap P_\i$. The UC action, on the other hand, removes all material in $P_\i - P$, including the minimal amount of material (e.g., in volumetric terms) in $P \cap P_\i$ that enables the removability of $P_\i - P$ with $T$. The next workpiece shape (output state) $P_\o \subset \R^3$ is contained within the input state in both cases (i.e., $P_\o \subseteq P_\i)$, however, the removed material is strictly outside the target shape (i.e., $(P_\i - P_\o) \cap P = \emptyset$) in the case of OC---meaning that there is no ``collateral damage''---while minimal collateral damage is tolerated in the case of UC such that all of $P^c$ is removed (i.e., $P_\o \subseteq P$).%
\footnote{The superscript $c$ means regularized set complement.}

\begin{figure*}
	\centering
	\includegraphics[width=\linewidth]{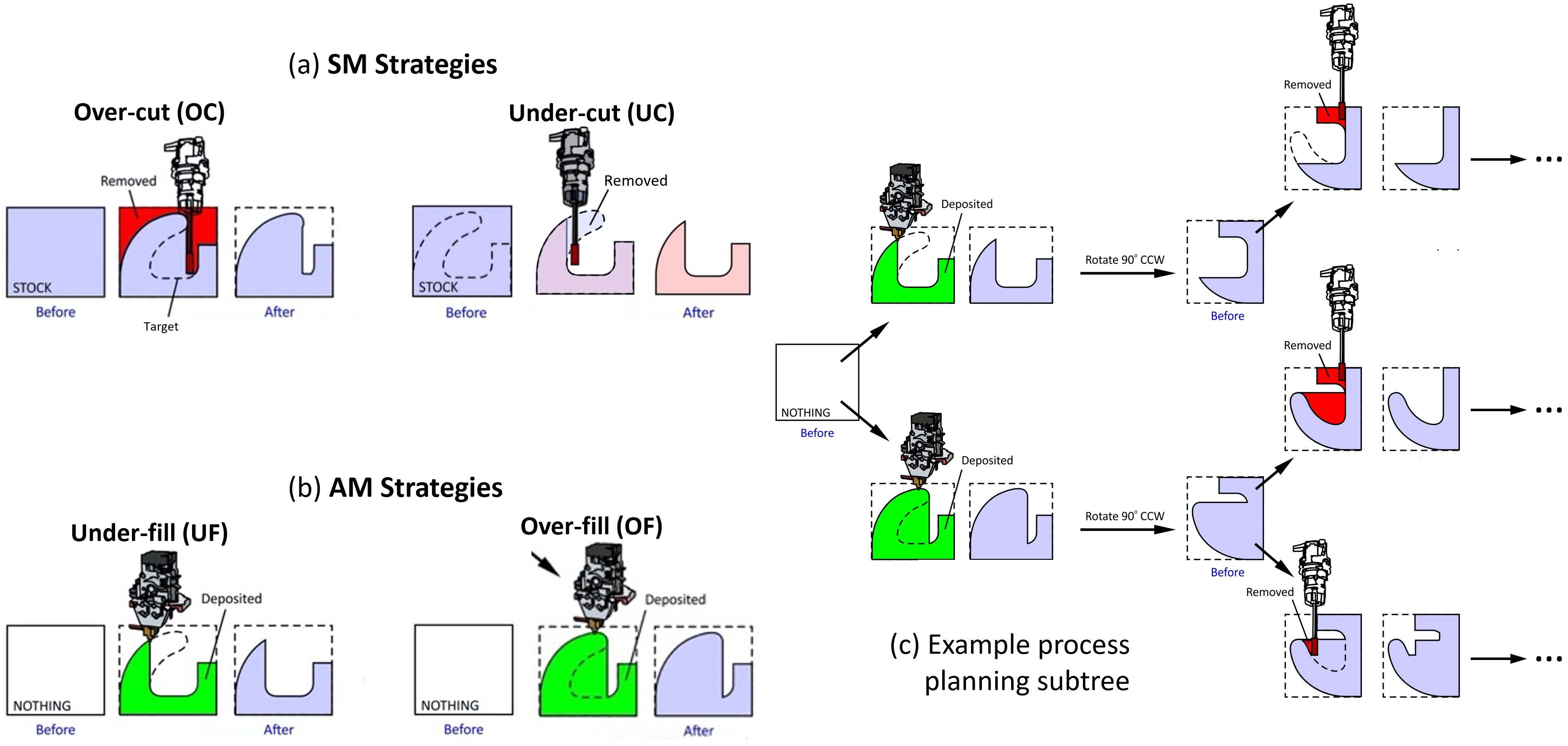}
	\caption{(a) The valid SM actions: OC and UC. (b) The valid AM actions: UF and OF. (c) An example (illustrative) process planning subtree.}
	\label{fig_cartoon}
\end{figure*}

Similarly, for AM, `valid' conservative and aggressive actions are defined, respectively, as follows:
\begin{itemize}
	\item An under-fill (UF) action is a one-shot material deposition operation that deposits as much deficit material as possible onto the given workpiece state without adding any region that is outside the target state.
	\item An over-fill (OF) action is a one-shot material deposition operation that deposits the entire deficit material onto the given workpiece state, regardless of whether it requires deposition of additional regions that are outside the target state.
\end{itemize}
The ``deficit'' material in both definitions refers to material that is inside the target state but outside the current state. The deposition of deficit material and/or additional regions are constrained by accessibility and support requirements (i.e., no overhang allowed). More precisely, for a given target shape $P \subset \R^3$, current part shape (input state) $P_\i \subset \R^3$, and tool shape $T\subset \R^3$, the UF action deposits as much of the deficit material (i.e., maximal subset of $P - P_\i$) that is accessible by $T$, as well as self-supporting for a given build direction, without having to deposit anything in $P^c - P_\i$. The OF action, on the other hand, deposits all material in $P - P_\i$, including as little material in $P^c$ as possible in order to enable depositability of $P - P_\i$ with $T$. The next workpiece shape (output state) $P_\o \subset \R^3$ contains the input state in both cases (i.e., $P_\i \subseteq P_\o)$, however, the deposited material is strictly inside the target shape (i.e., $(P_\o - P_\i) \subseteq P$) in the case of UF---meaning that there is no ``sacrificial material''---while minimal sacrificial material is tolerated in the case of OF such that all of $P$ is deposited (i.e., $P \subseteq P_\o$).

Figure \ref{fig_cartoon} (a, b) illustrates the 4 types of actions for the special case in which $P$ is a raw stock in the case of SM and nothing in the case of AM, leading to multiple branches in the search tree, some of which are illustrated in Fig. \ref{fig_cartoon} (c).

Despite the apparent duality between AM and SM actions, leading to a largely symmetric mathematical formulation in terms of set union and difference (i.e., intersection with complement), thanks to De Morgan's laws, there are a few symmetry breaking conditions.

First, the accessibility conditions imposed by the evolving part geometry are different for AM and SM. Although the addition of material during a given AM action gradually enlarges the $\conf-$space obstacle and limits accessibility, it is reasonable to assume that the motion for every AM action is ordered in a layer-by-layer fashion along the build direction and the print-head is always located above the previously deposited layer in the same action. Hence, the accessibility analysis based on the {\it input} state $P_\i$ of the workpiece remains valid throughout the AM action. We can make similar assumptions for SM motion ordering and the tool assembly so the accessibility is improved gradually during the motions. Hence, the $\conf-$space obstacle can be calculated based on the outcome of the SM action, i.e., the {\it output} state $P_\o$ of the workpiece, which is unknown to begin with. Hence, the SM action can only be defined as a self-referencing set equation that can be solved by fixed point iteration.

Second, for AM technologies that are commonly used in HM (e.g., DED laser cladding), there is an additional support constraint; namely, overhangs cannot be tolerated. For every deposited point in the same AM action, every other point beneath it must be present either in the input state or deposited in the same action. For the latter, the layer-by-layer ordering mentioned above ensures that the point in question is supported.

\section{Geometric Operations} \label{sec_geom}

In this section, after presenting the preliminary assumptions (Section \ref{sec_prelim}), we discuss the geometric operations to compute regions of 3D workspace that are accessible/inaccessible for both AM and SM (Section \ref{sec_access}), make the inaccessible regions accessible with minimal-collision collateral damage for SM (Section \ref{sec_collat}), and represent self-supporting min/max bounds for AM (Section \ref{sec_supp}).

\subsection{Preliminaries} \label{sec_prelim}

We consider the state space (which contains the initial, intermediate, and target states) to be the collection of all bounded solids or `r-sets', i.e., compact regular semianalytic sets in $\R^3$ \cite{Requicha1980representations}. For each action, $P_\i, P_\o \subset \R^3$ represent the input and output states, respectively, while the target state is given by $P \subset \R^3$. The tool assemblies (hereafter referred to as tools, for short) are also assumed to be solids. All operations on solids will be assumed to be regularized \cite{Requicha1978mathematical}.

Each AM or SM tool $T = K \cup H$ consists of two portion; namely, a {\it passive} portion $H$ that represents a large moving object that must avoid collision with stationary objects (i.e., the part and fixtures, if applicable); and an {\it active} portion $K$ that represents the unit of deposition/removal, i.e., a minimum manufacturable neighborhood (MMN) whose sweep along the motion characterizes the deposited/removed material for each AM/SM action, respectively. For SM tools, the passive portion often consists of milling/turning tool shank, holder, spindle, and other head components, while the active portion is the sharp edge of a cutter or tool insert. For AM tools, the passive portion can be a powder/wire-feeder, droplet jetting nozzle, and other print-head components, while the active portion may be defined in terms of a representative solidified droplet or melt-pool shape. Given the recent advancements in the resolution of AM processes, we ignore the minimum feature size requirement for the AM actions and assume $T \approx H$ for accessibility analysis, while the minimum feature size/shape for SM is encoded into the accessibility analysis, as we note in Section \ref{sec_access}.

Note that the motion is relative, hence we assume, without loss of generality, that the workpiece is fixed and the tool moves, while in reality it might be the opposite or both may move. To simplify the analyses, we assume that each tool is oriented with respect to the workpiece by a fixed rotation $R \in \SO{3}$ and the motion DOF after that is restricted to $3-$axis translations $\bt \in \R^3$. Note that for AM, assuming the print-head always faces downwards, the relative tool rotation by $R$ implies changing the build direction by applying an opposite rotation $R^{-1}$ to the part (e.g., refixturing), hence the gravity direction changes from vertical $(0, 0, -1)$ to $R^{-1}(0, 0, -1)$ for support analysis. For SM, on the other hand, it does not matter how we interpret the rotation, i.e., as rotating the head, the workpiece, or a combination of the two on a high-axis CNC machine, as long as the rotation remains fixed throughout each action. The choice of an optimal direction for both AM and SM will be incorporated into our search algorithm in section \ref{sec_search} by exploring a subset of the possible relative rotations of a given workpiece at each step of the process plan.

In each section, in addition to the set-theoretic formulation, we present the numerical algorithms to compute a Cartesian volumetric enumeration (i.e., voxelization) of the pointsets, although other representation schemes (e.g., B-reps or mesh) can be used as long as the Boolean and Minkowski operations are supported. More specifically, we present algorithms for computing discretized indicator functions (i.e., 3D binary-valued voxel arrays) and will denote such arrays using characteristic functions enclosed in brackets (e.g., $[\indic_{X}]$); real-valued arrays resulting from discrete convolutions will be denoted similarly, e.g., $[ \indic_{X} \ast \indic_{Y}] = \textsc{Conv} \big( [\indic_{X}], [\indic_{Y}] \big)$, were $\textsc{Conv}$ is FFT-based discrete convolution that can be GPU-accelerated.

\subsection{Accessibility Analysis} \label{sec_access}

For a given (AM or SM) tool $T \subset \R^3$ and an obstacle pointset $O \subset \R^3$ with which it must avoid collisions,%
\footnote{To simplify the notation, we will assume that $O := P_\i$ (for AM) or $P_\o$ (for SM), but one can incorporate additional objects in the environment (e.g., fixtures) in the obstacle set.}
we define the translational $\conf$-space obstacle $\mathcal{O}(O, T, R) \subseteq \R^3$ to be the set of translations $\bt \in \R^3$ of the rotated tool $RT$ that collide with the obstacle pointset $O$:
\begin{equation}
	\mathcal{O}(O, T, R) \triangleq \big\{\bt \in \R^3 ~|~ O \cap (RT + \bt) \neq \emptyset \big\}.
\end{equation}
With this definition in place, the set of non-colliding (i.e., {\it accessible}) translations is $\mathcal{O}(O, T, R)^c$. The translational $\conf-$space can be computed via a Minkowski sum:%
\footnote{Some of these equalities do not strictly hold upon regularization when the $\conf-$space obstacle has lower-dimensional features, but we shall ignore such degenerate cases. The drawback is that the technique may not be robust in such cases, e.g., when the target shape has a cylindrical hole that has exactly the same diameter as a milling tool.}
\begin{equation}
	\mathcal{O}(O, T, R) \eqae O \oplus (-RT). \label{eq_cobs}
\end{equation}
Given a minimum manufacturable neighborhood (MMN) $K \subset \R^3$, the set of points in the 3D workspace that can be touched (i.e., deposited or removed) by sweeping the MMN along an accessible motion is:
\begin{align}
	A(O, T, K, R) &\triangleq \mathcal{O}(O, T, R)^c \oplus (RK) \\
	&= \big( O \oplus (-RT) \big)^c \oplus (RK) \\
	&= \big( O^c \ominus (-RT) \big) \oplus (RK), \label{eq_access}
\end{align}
where the second equality results from the duality between Minkowski sum $\oplus$ and difference $\ominus$ as a result of De Morgan's laws \cite{Serra1983image}. We call this pointset ``accessible'' by the rotated tool $RT$ with respect to the obstacle pointset $O$. The inaccessible region, on the other hand, is simply defined by the remaining portion of $O^c$:
\begin{equation}
	I(O, T, K, R) \triangleq O^c - A(O, T, K, R). \label{eq_inaccess}
\end{equation}

\begin{figure}
	\centering
	\includegraphics[width=\linewidth]{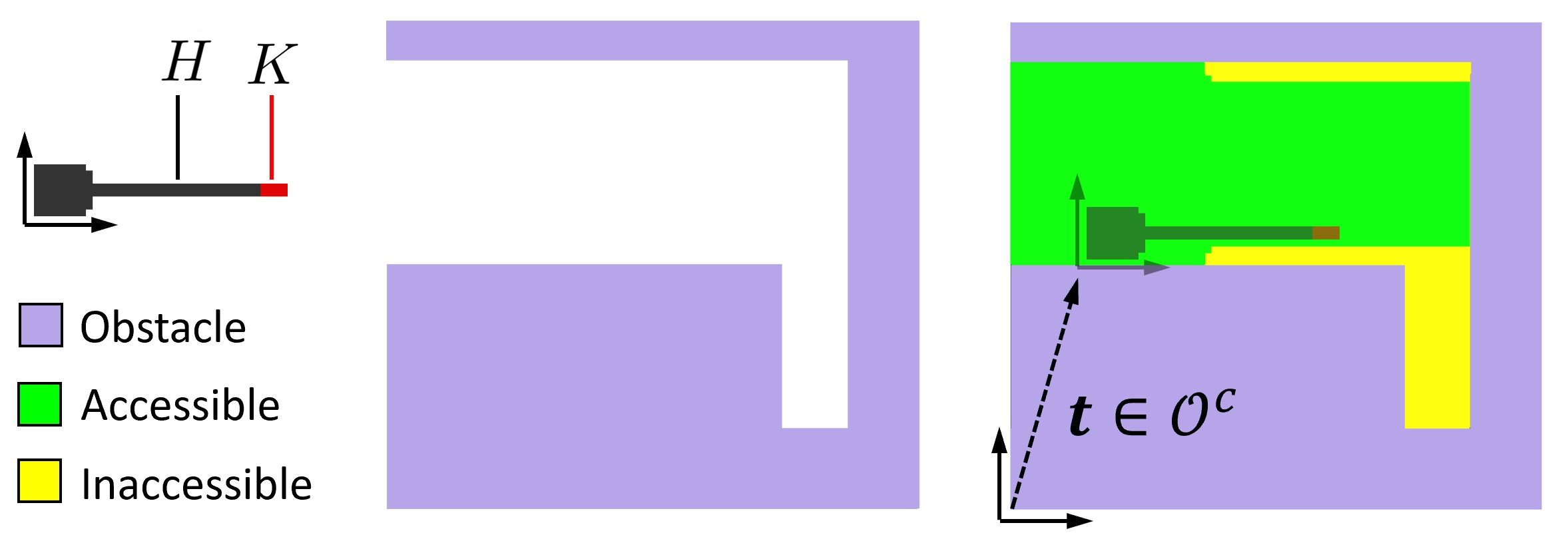}
	\caption{A depiction of the milling tool in its local coordinate system (left) along with the target part shape (middle), the accessible region $A$, and the inaccessible region $I$ for identity rotation $R := \id$.}
	\label{fig_access}
\end{figure}
Figure \ref{fig_access} illustrates the accessible $A$ and inaccessible $I$ regions for a target shape $P$, used (initially) as the obstacle set $O := P$, and an SM tool $T = H \cup K$ in 2D.%
\footnote{Throughout the paper, we use simple 2D examples for illustrative purposes. The simplicity of such shapes (e.g., straight line boundaries) is incidental, and does not imply any such limitations in practice, as we show in Section \ref{sec_results} for 3D parts and tools of arbitrary shapes.}

To compute the $\conf-$space obstacles efficiently, we note the correspondence between Minkowski and convolution algebras \cite{Guibas1983kinetic,Lysenko2010group}. If we represent a pointset $X \subseteq \R^3$ implicitly via its indicator (i.e., characteristic) function:
\begin{align}
	\indic_{X}(\bx) \triangleq \left\{
	\begin{array}{ll}
		1 & \text{if}~ \bx \in X,\\
		0 & \text{otherwise,}
	\end{array} \right.
\end{align}
then we may represent the Minkowski sum of two pointsets implicitly as the convolution of their indicator functions:
\begin{equation}
	\indic_{X \oplus Y}(\bx) > 0 ~\overset{\mathsf{ae}}{\rightleftharpoons}~ \left( \indic_{X} * \indic_{Y} \right)(\bx) > 0,
\end{equation}
which can be computed efficiently via fast Fourier transform (FFT) when the indicator functions are sampled over a uniform Cartesian grid (i.e., the pointsets are voxelized).

The translational $\conf-$space obstacle in \eq{eq_cobs} is thus computable as follows:
\begin{equation}
	\indic_{\mathcal{O}(O, T, R)}(\bx) \eqae \sign (\indic_O \ast \tilde{\indic}_{RT})(\bx), \label{eq_cobs_imp}
\end{equation}
where $\tilde{\indic}_{X}(\bx) = \indic_{X}(-\bx)$ represents a reflection with respect to the origin.
The accessible translations are computable by a logical negation: $\indic_{\mathcal{O}(O, T, R)^c} (\bx) = \neg \indic_{\mathcal{O}(O, T, R)} (\bx)$.
Hence, the accessible region in \eq{eq_access} can be computed as:
\begin{align}
	\indic_{A(O, T, K, R)}(\bx) &\eqae \sign (\neg \indic_{\mathcal{O}(O, T, R)} \ast \indic_{RK}) (\bx) \nonumber \\
	&= \sign \Big(\big(\neg \sign (\indic_O \ast \tilde{\indic}_{RT}) \big) \ast \indic_{RK} \Big) (\bx). \label{eq_access_imp}
\end{align}

The inaccessible region in \eq{eq_inaccess} is thus computed as:
\begin{equation}
	\indic_{I(O, T, K, R)} = \neg\indic_O(\bx) - \indic_{A(O, T, K, R)}(\bx), \label{eq_inaccess_imp}
\end{equation}
noting that $\indic_{A(O, T, K, R)}(\bx) \leq \indic_{O^c}(\bx) = \neg\indic_O(\bx)$ because $A(O, T, K, R) \subseteq O^c$. 

Algorithms \ref{alg:A} and \ref{alg:I} describe how the accessible and inaccessible regions are computed, respectively, for a given (i.e., explicitly known) obstacle pointset. In Sections \ref{sec_OC} and \ref{sec_UC}, we show how these algorithms can be used to compute such regions when the obstacle pointsets are not explicitly known, but are implicitly defined through recursion.

\begin{algorithm}[hbt!]
	\caption{Computing accessible region}
	\begin{algorithmic}
		\Procedure{Accessible}{$[\indic_O]$, $[\indic_T]$, $[ \indic_K], R$}
		\State $[\indic_{RK}] \gets \textsc{Rotate} \big( [\indic_K], R \big)$
		\State $[\indic_{RT}] \gets \textsc{Rotate} \big( [\indic_T], R \big)$
		\State $[\tilde{\indic}_{RT}] \gets \textsc{Reflect} \big( [\indic_{RT}] \big)$
		\State $[\indic_{\mathcal{O}}] \gets \textsc{Conv} \big( [\indic_O], [\tilde{\indic}_{RT}] \big) > 0$ \Comment{Elementwise sign}
		\State $[\indic_A] \gets \textsc{Conv} \big( \neg[\indic_{\mathcal{O}}], [\indic_K] \big) > 0$ \Comment{Elementwise sign}
		\State \Return $[\indic_A]$
		\EndProcedure
	\end{algorithmic} \label{alg:A}
\end{algorithm}

\begin{algorithm}[hbt!]
	\caption{Computing inaccessible region}
	\begin{algorithmic}
		\Procedure{Inaccessible}{$[\indic_O]$, $[\indic_T]$, $[ \indic_K], R$}
		\State \Return $\neg[\indic_O] - \textsc{Accessible}\big([\indic_O]$, $[\indic_T]$, $[ \indic_K], R\big)$
		\EndProcedure
	\end{algorithmic} \label{alg:I}
\end{algorithm}

\subsection{Collateral Analysis} \label{sec_collat}

Let $\bx \in I(O, T, K, R)$ be a point (called a `query' point) in the inaccessible region with respect to an obstacle pointset $O$ and an SM tool $T = H \cup K$. Let us consider all possible ways in which any point in the cutter $\bk \in K$ that can remove material (called a `sharp' point \cite{Mirzendehdel2020topology}) can be brought into contact with the query point. Our goal is to identify the sharp point, using which the collision between the tool in this configuration and the obstacle set is minimized in some way (e.g., in terms of volume). The union of such regions over all inaccessible query points, intersected with the target shape $P$, characterizes minimal ``collateral damage'' in the sense that it is as little material as we need to remove inside the target shape to make all those query points accessible. This material has to be brought back later by an AM tool.

Let $\rho^{}_\text{CMF} (\bx, \bk; P, T, R)$ stand for the bi-variate collision measure field (CMF) between the target shape $P$ and the SM tool $T$ rotated by $R$ and translated in such a way the the sharp point $\bk \in K$ is brought into contact with the query point $\bx \in I(O, T, K, R)$. The CMF can thus be computed as an intersection measure:
\begin{equation}
	\rho^{}_\text{CMF} (\bx, \bk; P, T, R) \triangleq \mu^3 \Big[ P \cap \big( R (T - \mathbf{k}) + \bx \big)\Big], \label{eq_rho}
\end{equation}
where $\mu^3[\cdot]$ denotes the Lebesgue $3-$measure (i.e., volume).
The sharp point that results in minimal CMF for a given query point $\bx \in I(O, T, K, R)$ is thus given by:
\begin{equation}
	\bk_{\min}(\bx; P, T, K, R) \triangleq \underset{\mathbf{k} \in K}{\arg\min} ~\rho^{}_\text{CMF} (\bx, \bk; P, T, R), \label{eq_kmin}
\end{equation}
It is possible that the result is not unique, i.e., more than one sharp point exhibit the same minimal collision volume. In that case, we may assume an arbitrary total ordering $(K, \prec)$ to prioritize between $\bk_{\min} \prec \bk'_{\min}$ if both result in a minimal CMF.

Note that $R (T - \mathbf{k}) + \bx$ in \eq{eq_rho} indicates translating the tool to bring the sharp point to the origin, applying a rotation $R$, and translating the tool once again to take the sharp point from the origin to the query point, where its intersection with the target shape is measured. The total translation is thus obtained as:
\begin{equation}
	\bt_{\min}(\bx; P, T, K, R) \triangleq \bx -R\bk_{\min}(\bx; P, T, K, R).
\end{equation}
The minimal-collision collateral damage for a single query point can be obtained by intersecting the moved tool (using the rotation $R$ and the above translation) with the target shape $P$. The union of all such intersections over all inaccessible query points yields:
\begin{align}
	C(O, P, T, K, R) &= P \cap \!\!\!\! \!\!\!\! \bigcup_{\bx \in I(O, T, K, R)} \!\!\!\! \!\!\!\! \big(R T + \bt_{\min}(\bx) \big) \nonumber \\
	&= P \cap \!\!\!\! \!\!\!\!  \bigcup_{\bx \in I(O, T, K, R)} \!\!\!\! \!\!\!\! R\big(T - \bk_{\min}(\bx)\big) + \bx. \label{eq_C}
\end{align}
where the arguments in $\bt_{\min}(\bx) = \bt_{\min}(\bx; P, T, K, R)$ are dropped to simplify the notation, and the common intersection is factored out of the union.

For computational purposes, it is more practical to invert the dictionary look-up from query points to the corresponding minimal-collision sharp points in \eq{eq_kmin}. For a given sharp point $\bk \in K$, let $Q(\bk; O, P, T, K, R)$ stand for the subset of all inaccessible points $\bx \in I(O, T, K, R)$ for which $\bk = \bk_{\min}(\bx; P, T, K, R)$. This condition can be replaced with the following equality test:
\begin{equation}
	\rho^{}_\text{CMF} (\bx, \bk; O, T, R) == \min_{\bk \in K} \rho^{}_\text{CMF} (\bx, \bk; O, T, R),
\end{equation}
only if the minimal-collision sharp point for every query point is unique, which may not be the case. To mitigate the non-uniqueness, the above condition can be conjuncted with an additional condition that the query point was not already accounted for in a higher-priority sharp point with the same CMF, i.e., by excluding $Q(\bk)$ from $Q(\bk')$ if $\bk \prec \bk'$, to avoid double-counting.

Using the inverted definition, the pointset in \eq{eq_C} can be alternatively computed as follows:
\begin{align}
	C(O, P, T, K, R) &= P \cap \bigcup_{\bk \in K} \!\! \Big( R(T - \bk) + \!\!\! \bigcup_{\bx \in Q(\bk)} \!\!\! \bx \Big) \nonumber\\
	&= P \cap \bigcup_{\bk \in K} \!\! \Big( R(T - \bk) \oplus Q(\bk) \Big), \label{eq_C_mink}
\end{align}
where the arguments in $Q(\bk) = Q(\bk; O, P, T, K, R)$ are once again dropped to simplify the notation.

To compute the CMF defined in \eq{eq_rho} efficiently, we can use the convolution operator as follows:
\begin{equation}
	\rho^{}_\text{CMF} (\bx, \bk; P, T, R) \eqae \big(\indic_P \ast \indic_{R (T - \mathbf{k})} \big) (\bx),
\end{equation}
The Minkowski sum, union, and intersection in \eq{eq_C_mink} can also be implicitly computed as a convolution, summation, and multiplication, respectively:
\begin{equation}
	\indic_C (\bx) \eqae \indic_P (\bx) \cdot \sign \left( \sum_{\bk \in K} \indic_{R(T - \bk)} \ast \indic_{Q(\bk)}(\bx) \right),
\end{equation}
where we use $C = C(O, P, T, K, R)$ to simplify notation.

Algorithm \ref{alg:C} describes the computation of the minimal-collision collateral damage.
The discretization of indicator functions and CMF, denoted by the brackets $[\cdot]$, is with respect to the query point $\bx \in \R^3$ indexed over a uniform Cartesian grid, while the sharp points $\bk \in (K, \prec)$ are sampled and ordered for a sequential for-loop. For all practical purposes, sampling only on the exposed sharp edges of the boundary $\partial K$ is sufficient. The algorithm uses the subroutine $\textsc{Inaccessible}([\indic_O], [\indic_{T}], [\indic_{K}], R)$ from Algorithm \ref{alg:I}.

\begin{algorithm}[hbt!]
	\caption{Calculating collateral region}\label{alg:OC}
	\begin{algorithmic}
		\Procedure{Collateral}{$[\indic_O], [\indic_P], [\indic_{T}], [\indic_{K}], R$}
		\State $[\indic_C] \gets [\indic_{\emptyset}]$
		\State $[\indic_{RT}] \gets \textsc{Rotate} \big([\indic_{T}], R \big)$ 
		\For{$\bk$ in $K$}
		\State $[\indic_{R(T - \bk)}] \gets \textsc{Translate} \big([\indic_{RT}], -R\bk \big)$ 
		\State $[\rho^{}_{\text{CMF}}(\bk)] \gets \textsc{Conv} \big( [\indic_P], [\indic_{R(T - \bk)}] \big)$ 
		\EndFor
		\State $[\rho^{\min}_{\text{CMF}}] \gets \min_{\mathbf{k} \in K} [\rho^{}_{\text{CMF}}(\bk)]$ 
		\State $[\indic_I] \gets \textsc{Inaccessible} \left([\indic_O], [\indic_{T}], [\indic_{K}], R \right)$
		\For{$\bk \in K$}
		\State $[\indic_{Q(\bk)}] \gets \Big( [\rho^{}_{\text{CMF}}] == \left[ \rho^{\min}_{\text{CMF}} \right] \Big) \cdot [\indic_I]$
		\State $[\indic_C] \gets  \Big( [\indic_C] + \textsc{Conv} \big([\indic_{Q(\bk)}], [\indic_{R(T-\bk)}] \big) \Big) > 0$ 
		\State $[\indic_I] \gets [\indic_I] - [\indic_{Q(\bk)}]$
		\Comment{Avoids double-counting}
		\EndFor
		\State $[\indic_C] \gets [\indic_C] \cdot [\indic_P]$
		\State \textbf{return} $[\indic_C]$
		\EndProcedure
	\end{algorithmic} \label{alg:C}
\end{algorithm}

\begin{figure}
	\centering
	\includegraphics[width=\linewidth]{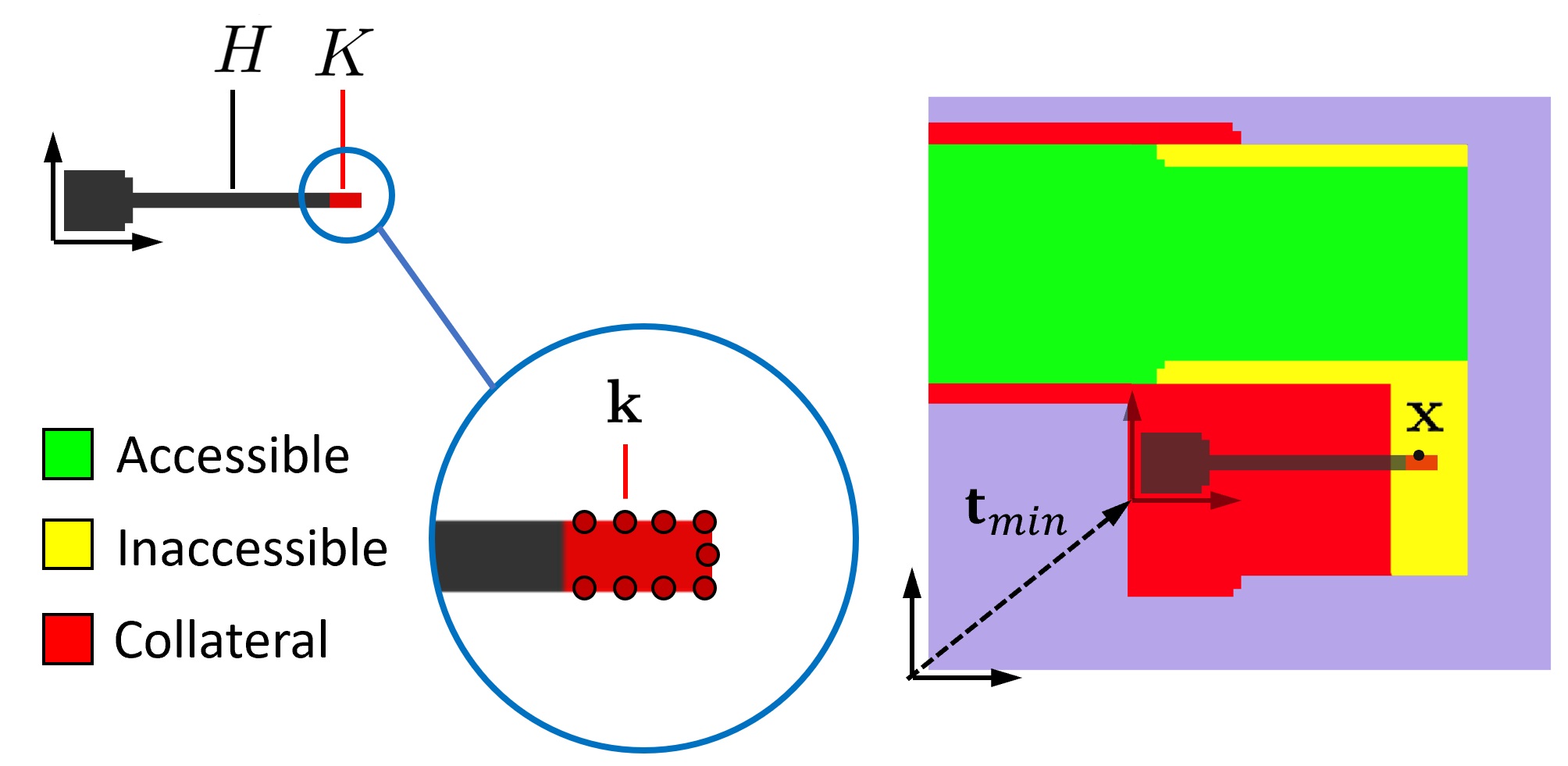}
	\caption{A depiction of the various sharp points $\bk \in K$ along the cutter edge which may be brought in contact with the query point $\bx \in O^c$, along with a depiction of the accessible $A$, inaccessible $I$, and minimal-collision collateral regions $C$.}
	\label{fig_sharp}
\end{figure}

Figure \ref{fig_sharp} illustrates the pointsets $A$, $I$, and $C$ for an example target part $P$ and SM tool $T = H \cup K$ in 2D, with a sampling of sharp points over the boundary of $K$.

\subsection{Support Analysis} \label{sec_supp}

To simplify the accessibility analysis, we assume that the AM tool cannot tolerate overhangs, i.e., the slightest deviation from a vertical wall would create problems. While many AM processes can handle overhang angles up to a certain value (e.g., $45^\circ$ is a commonly used threshold), the ones that are most commonly used in HM scenarios (e.g., DED/DMD for laser-based cladding) require fairly straight walls. This assumption makes it simpler and faster to compute well-defined self-supporting regions that are either maximally contained within or minimally contain the target shape, as depicted in Fig. \ref{fig_cartoon} (b).%
\footnote{If overhangs below a certain nonzero angle are tolerated, there is no straightforward way to define such max/min pointset uniquely.}

Let $\bx \in \R^3$ be a query point for which we aim to perform support analysis against a target shape $P$.

Let us first consider the set of self-supporting points $U(P, R) \subset \R^3$ that are maximally contained within the target shape $P$. If $\bx \in P$, then it can be printed without extra support material (i.e., material outside $P$) if and only if a straight half-line drawn downwards (i.e., along gravity) from the query point remains completely inside $P$ before reaching the build plate. This condition can be stated as:
\begin{equation}
	\bx \in U(P, R) ~\rightleftharpoons~ \mu^1 \big[P \cap (\bx + RL) \big] = (\bx-\bx_{\mathrm{B}}) \cdot (R\be_z), \label{eq_halfline_min}
\end{equation}
where $\be_z = (0, 0, 1)$ is the unit vector along the $z-$axis, and the build direction is given by the rotation $R$ of the AM tool, i.e., rotation $R^{-1}$ of the workpiece for an AM tool facing downwards along the negative $z-$axis. If our frame of reference is the workpiece, we can imagine the direction of gravity changing to $-R \be_z$. $L \triangleq \{\alpha \be_z ~|~ \alpha \leq 0 \}$ is an infinite half-line along the negative $z-$axis, $\bx_\mathrm{B} \in \R^3$ is any point on the surface of the build plate, and $\mu^1[\cdot]$ is the Lebesgue $1-$measure (i.e., length). The rotated and displaced half-line $\bx + RL$ stems out of the query point along the direction of gravity. If the length of its portion falling inside the pointset $P$ is the same as the height of the query point above the build plate along the build direction, then the query point is included in $U(P, R)$. In fact, every point on the line segment $P \cap (\bx + RL)$, including the query point, can be printed without the need for extra support. On the other hand, if any portion of the segment falls external to the target shape, the length of the external segment $P^c - (\bx + RL)$ will be smaller, hence the external segment must be printed with excess material to support the internal segment $P \cap (\bx + RL)$, including the query point.

Let us next consider the set of self-supporting points $V(P, R) \subset \R^3$ that minimally contain the target shape $P$. If $\bx \in P^c$, then it must be printed as extra support material (i.e., material outside $P$), if and only if a straight half-line drawn upwards (i.e., against gravity) from the query point intersects $P$. This condition can be stated as:
\begin{equation}
	\bx \in V(P, R) ~\rightleftharpoons~ \mu^1 \big[ P \cap (\bx + (-RL)) \big] > 0,
\end{equation}
meaning that there exists a segment of the half-line $P \cap (\bx + (-RL))$, stemming out of the query point against the direction of gravity, that intersects the target shape. Hence, the query point must be printed as excess material to support the said segment. On the other hand, if the half-line does not intersect with the target shape, the length of the segment $P \cap (\bx + (-RL))$ will be zero, hence the query point does not support anything and need not be printed. Note that $U(P, R) \subseteq P \subseteq V(P, R)$.

To compute the above max/min pointsets efficiently, we can replace the half-line $L$ with an infinitesimally thin cylinder $L_\epsilon \triangleq L \oplus [-\nicefrac{\epsilon}{2}, +\nicefrac{\epsilon}{2}]^3$ where $\epsilon \to 0^+$, hence replace the $1-$measure in the above equations with a $3-$measure:
\begin{align}
	\bx \in U(P, R) &~\overset{\mathsf{ae}}{\rightleftharpoons}~ \lim_{\epsilon \to 0^+} \frac{1}{\epsilon^2} \mu^3 \big[ P \cap (\bx + RL_\epsilon) \big] = h(\bx; R), \\
	\bx \in V(P, R) &~\overset{\mathsf{ae}}{\rightleftharpoons}~ \lim_{\epsilon \to 0^+} \frac{1}{\epsilon^2} \mu^3 \big[ P \cap (\bx + (-RL_\epsilon)) \big] > 0,
\end{align}
where $h(\bx; R)\triangleq (\bx-\bx_{\mathrm{B}}) \cdot (R\be_z)$ is the height field for the build direction specified by $R$. The intersection measures can, once again, be computed via convolutions:
\begin{align}
	\indic_{U(P, R)} (\bx) &\eqae \Big[ \lim_{\epsilon \to 0^+} \frac{1}{\epsilon^2}(\indic_P \ast \tilde{\indic}_{RL_\epsilon}) (\bx) == h(\bx; R) \Big], \\
	\indic_{V(P, R)} (\bx) &\eqae \Big[ \lim_{\epsilon \to 0^+} \frac{1}{\epsilon^2}(\indic_P \ast \indic_{RL_\epsilon}) (\bx) > 0 \Big].
\end{align}
For discretized shapes over a uniform Cartesian grid of voxels of edge length $\epsilon > 0$, the above formulae can be approximated by discrete convolutions $\textsc{Conv}\big([\indic_P], [\tilde{\indic}_{RL_\epsilon}]\big)$ (Algorithm \ref{alg:U}) and $\textsc{Conv}\big([\indic_P], [\indic_{RL_\epsilon}]\big)$ (Algorithm \ref{alg:V}), respectively, in which the arrays $[\indic_{L_\epsilon}]$ and $[\tilde{\indic}_{L_\epsilon}]$ represent vertical half-columns of a single voxel thickness, extending downwards and upwards from the origin, respectively.

\begin{algorithm}[hbt!]
	\caption{Computing maximal self-supported region}
	\begin{algorithmic}
		\Procedure{MaxSelfSupp}{$[\indic_P], R, \bx_\mathrm{B}$}
		\State $[h] \gets \big([\bx]-\bx_{\mathrm{B}} \big) \cdot (R\be_z)$ \Comment{Discretized height field}
		\State $[\indic_{L_\epsilon}] \gets \textsc{CreateDHC}()$ \Comment{Downward half-column}
		\State $[\indic_{RL_\epsilon}] \gets \textsc{Rotate} \big([\indic_{L_\epsilon}], R\big)$
		\State $[\tilde{\indic}_{RL_\epsilon}] \gets \textsc{Reflect} \big([\indic_{RL_\epsilon}]\big)$
		\State \Return $\textsc{Conv} \big([\indic_P], [\tilde{\indic}_{RL_\epsilon}]\big) == [h]$ 
		\State \Comment{Tested up to some numerical error tolerance}
		\EndProcedure
	\end{algorithmic} \label{alg:U}
\end{algorithm}

\begin{algorithm}[hbt!]
	\caption{Computing minimal self-supported region}
	\begin{algorithmic}
		\Procedure{MinSelfSupp}{$[\indic_P], R, \bx_\mathrm{B}$}
		\State $[h] \gets \big([\bx]-\bx_{\mathrm{B}} \big) \cdot (R\be_z)$ \Comment{Discretized height field}
		\State $[\indic_{L_\epsilon}] \gets \textsc{CreateDHC}()$ \Comment{Downward half-column}
		\State $[\indic_{RL_\epsilon}] \gets \textsc{Rotate} \big([\indic_{L_\epsilon}], R\big)$
		\State \Return $\textsc{Conv} \big([\indic_P], [\indic_{RL_\epsilon}]\big) > 0$ 
		\EndProcedure
	\end{algorithmic} \label{alg:V}
\end{algorithm}

\begin{figure}
	\centering
	\includegraphics[width=\linewidth]{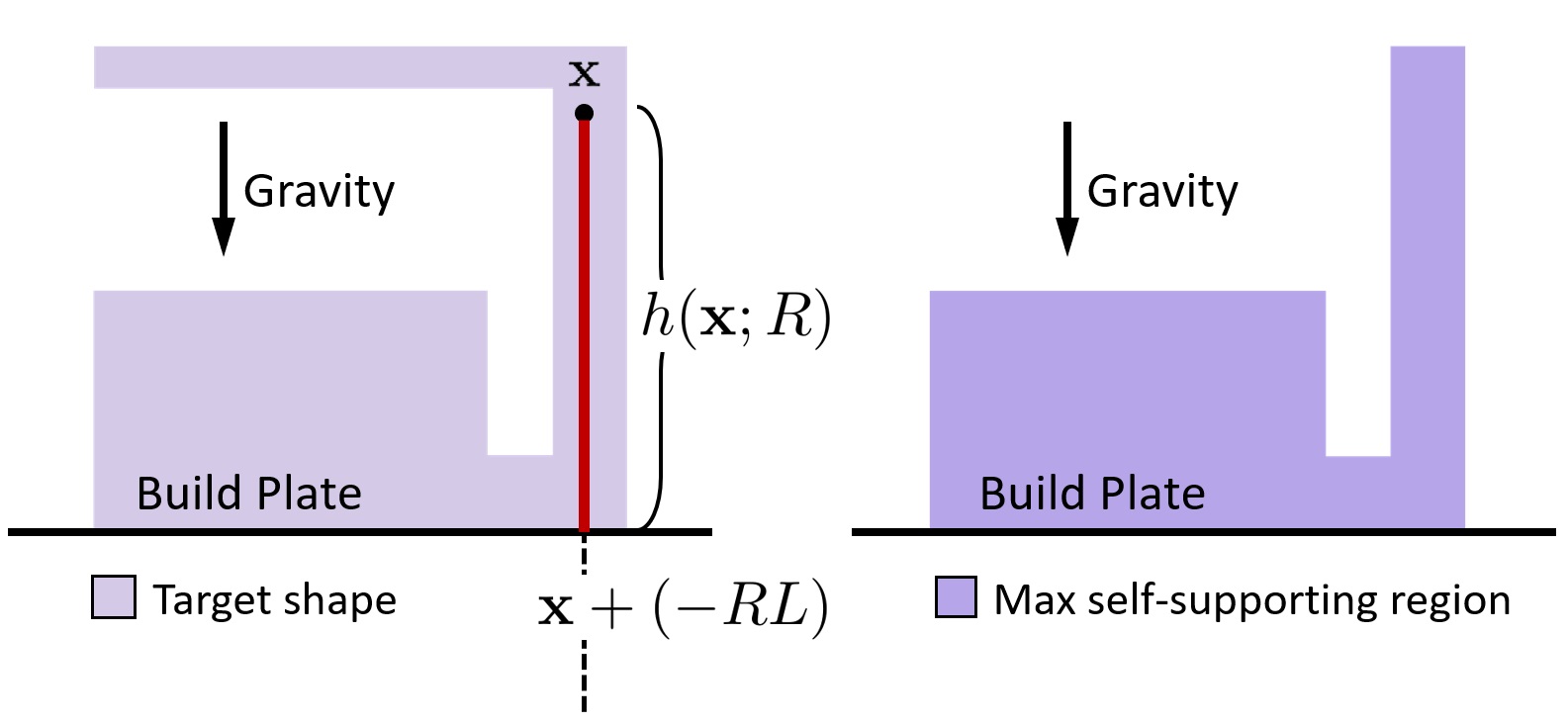}
	\caption{A depiction of the maximal self-supporting region $U$ contained within a target shape $P$, starting from an empty build plate.}
	\label{fig_maxsupp}
\end{figure}

\begin{figure}
	\centering
	\includegraphics[width=\linewidth]{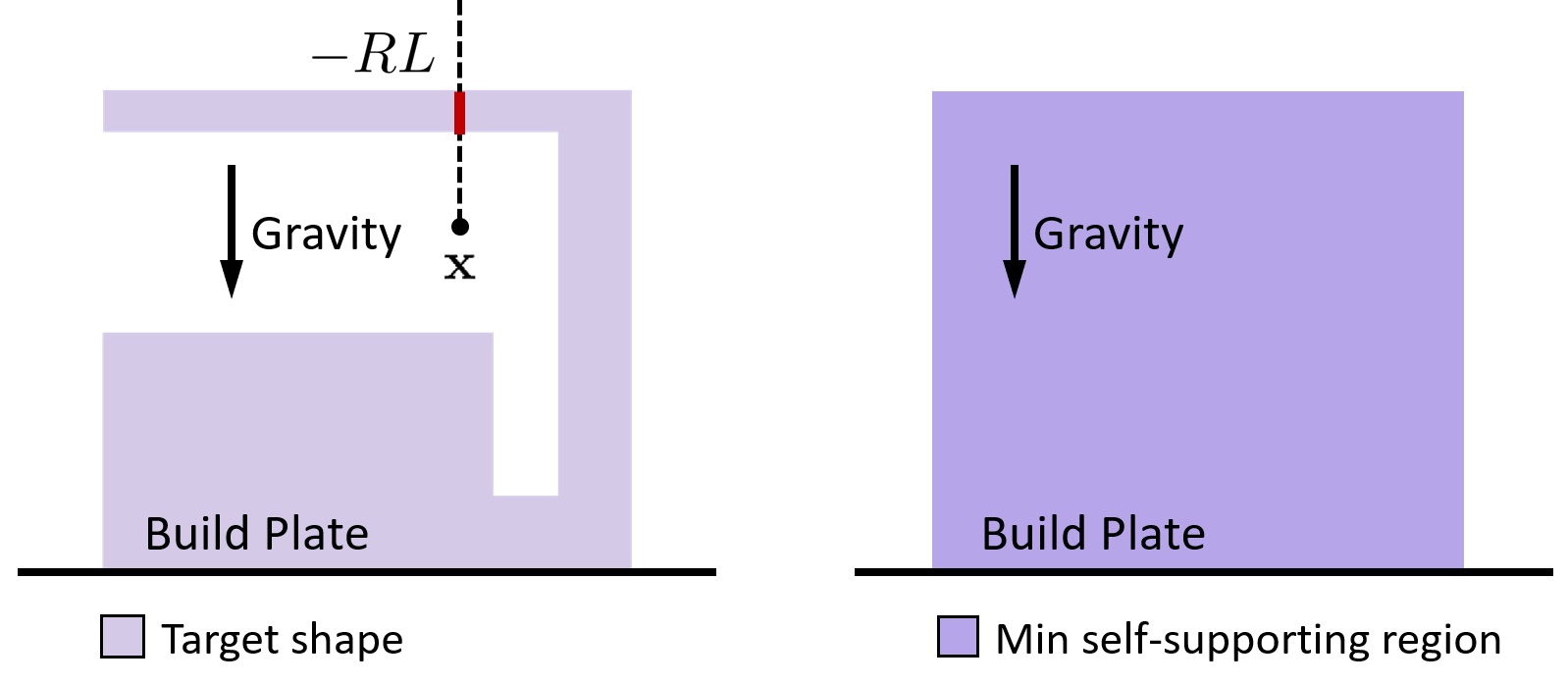}
	\caption{A depiction of the minimal self-supporting region $V$ containing a target shape $P$, starting from an empty build plate.}
	\label{fig_minsupp}
\end{figure}

Figures \ref{fig_maxsupp} and \ref{fig_minsupp} illustrate the max/min self-supporting regions, respectively, for a simple example in 2D, for upward build direction (i.e., $R := \id$).

\section{Manufacturing Actions} \label{sec_actions}

Having defined and implemented the basic building blocks (namely, accessibility, collateral, and support analyses) in the previous section, we define the SM (OC/UC) (Section \ref{sec_SM}) and AM (UF/OF) (Section \ref{sec_AM}) actions precisely to define the actions for the HM process planning search space.

\subsection{SM Actions} \label{sec_SM}

With the accessible, inaccessible, and collateral regions defined, we can precisely formulate the valid SM actions according to our conservative and aggressive policies; namely, under-cut (UC) and over-cut (OC), respectively.

While the accessibility analysis in Section \ref{sec_access} provides the accessible translations for an SM tool $T$ against a {\it fixed} obstacle pointset $O$, it cannot readily address the removability of material from a given workspace state $P_\i$, since it does not account for the evolution of the workpiece geometry and the resulting changes in accessibility throughout the continuous SM action. In other words, the resulting workpiece state after taking out the removable region $A(O, T, K, R)$, defined in \eq{eq_access} and \eq{eq_access_imp} with respect to accessibility and MMN, depends on the obstacle pointset $O := P_\o$, which, in turn, depends on the resulting workpiece state, leading to a self-referencing set equation. We will solve the equation recursively by fixed point iteration.

\begin{figure}
	\centering
	\includegraphics[width=\linewidth]{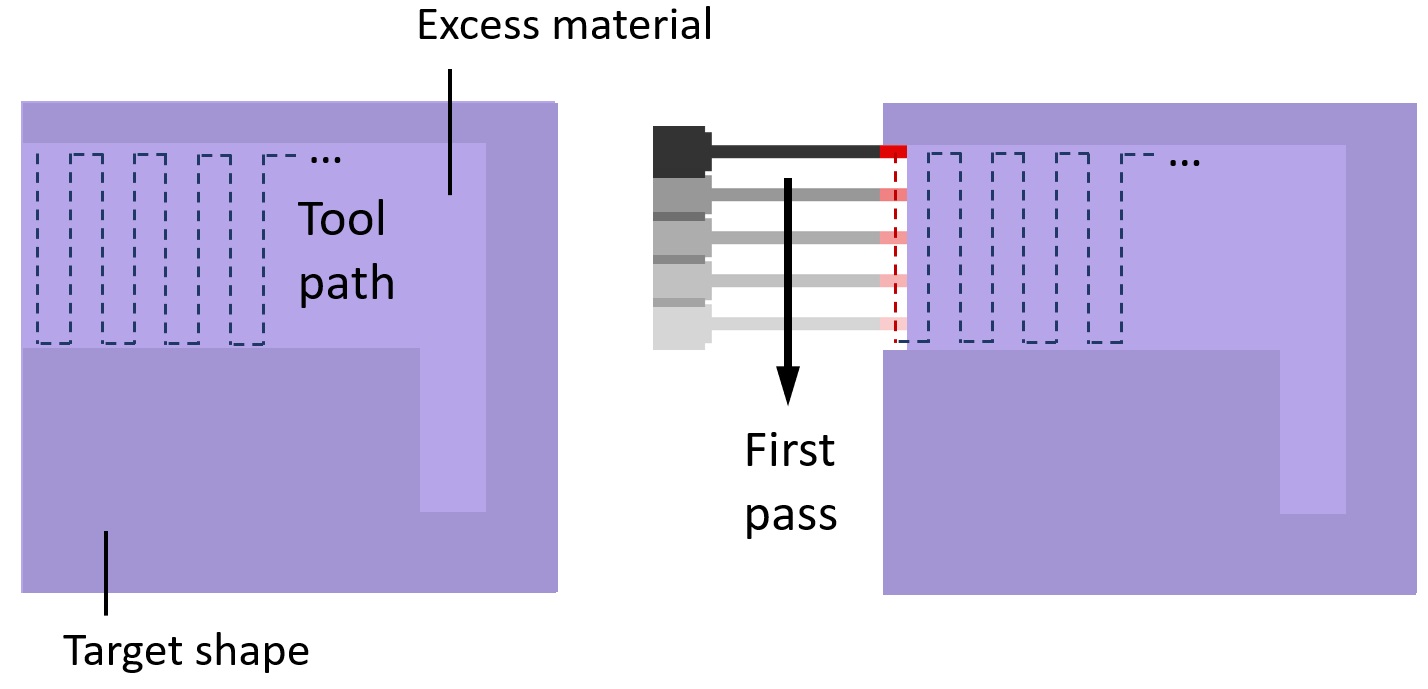}
	\caption{Given an ordering on the motion (e.g., a tool path), the obstacle set evolution can be accounted for incrementally, e.g., the obstacle for each vertical pass is the target shape minus the excess material removed in the previous vertical passes, which is expensive.}
	\label{fig_pass}
\end{figure}

To gain some intuition about this process, consider a typical $3-$axis milling process, the tool $T_\mathrm{SM}$ mills away material in multiple passes. At every pass, the accessibility changes as the $\conf-$obstacle is shaved off. One possible solution is to repeat the accessibility analysis at every pass to compute the allowable configurations of the tool for the next pass. Such computations would be prohibitively inefficient, because the depth of material removed in each pass cannot exceed that of the MMN (e.g., cutter depth), which may be quite small compared to the excess material region, as depicted in Fig. \ref{fig_pass}. Alternatively, we take a ``backward'' iteration approach, formalized by an iterative solution of a self-referencing equation. This approach converges more quickly since each pass removes material on the order of the length of the entire tool. We elaborate how it works for the OC and UC cases in detail below.

\subsubsection{Over-Cut Actions} \label{sec_OC}

As described at the beginning of section \ref{sec_revisit}, an OC action removes as much material from the region $P_\i - P$ as possible, while avoiding any material removal from $P \cap P_\i$. 

Let $P_\o$ represent the output of the OC action whose inputs are the target shape $P$, the current shape (before the action) $P_\i$, the SM tool $T = H \cup K$ with passive part $H$ (e.g., the holder) and active part $K$ (e.g., the cutter), oriented with a fixed rotation $R$. The output pointset must satisfy the following set equation:
\begin{equation}
	P_\o = P_\i - A(P_\o, T, K, R). \label{overcut_identity}
\end{equation}
This equation means that the output of the OC action is itself treated as the obstacle pointset $O := P_\o$, against which the accessible region $A(O, T, K, R)$ is computed, whose removal from the input state $P_\i$ results in the output state $P_\o$. Obviously, due to the complex algebraic form of the formula for $A(P_\o, T, K, R)$ in terms of two embedded Minkowski sums in \eq{eq_access} or convolutions in \eq{eq_access_imp}, the above equation cannot be solved by algebraic manipulation. But it can be solved recursively until its fixed point is found.

We begin the iterative process by an initial {\it underestimation} of the obstacle pointset $O_0 := P_\i \cap P$, noting that the OC policy, by definition, will result in an outcome $O = P_\o$ that contains $O_0$. In other words, $A_0 := A(O_0, T, K, R)$ is an {\it overestimation} of the accessible region, as it does not account for tool collisions with the inaccessible region $O - O_0$. The OC output is thus underestimated as $O_1 := P_\i - A_0$, providing a slightly better overestimation of the accessible region $A_1 := A(O_1, T, K, R)$, hence a slightly better underestimation of the OC output $O_2 := P_\i - A_1$, and so on. Proceeding with this iteration, we obtain a monotonically decreasing sequence of pointsets overestimating the accessible region, upper-bounded by the true region:
\begin{equation}
	A_0 \supseteq A_1 \supseteq \cdots \supseteq A = A(P_\o, T, K, R), \quad
\end{equation}
and a monotonically increasing sequence of pointsets underestimating the OC output (and the true obstacle set), lower-bounded by the true region:
\begin{equation}
	O_0 \subseteq O_1 \subseteq \cdots \subseteq O = \OC(P_\i, P, T, K, R),
\end{equation}
where $P_\o = \OC(P_\i, P, T, K, R)$ represents the converged solution for the OC output. Note that by construction, the fixed point of recursion (i.e., the pointset to which the iteration converges) obeys the identity in \eq{overcut_identity}.

Algorithm \ref{alg:OC} describes the recursive method used to compute the OC output. The algorithm uses the subroutine $\textsc{Accessible}([\indic_O], [\indic_{T}], [\indic_{K}], R)$ from Algorithm \ref{alg:A}.

\begin{algorithm}[hbt!]
	\caption{Computing OC output}
	\begin{algorithmic}
		\Procedure{OverCut}{$[\indic_P]$, $[\indic_{P_\i}]$, $[\indic_{T}]$, $[ \indic_{K}], R$}
		\State $[\indic_O] \gets [\indic_\emptyset]$
		\State $[\indic_{P_\o}] \gets [\indic_P]$
		\While{$[\indic_O] \neq [\indic_{P_\o}]$}
		\State $[\indic_O] \gets [\indic_{P_\o}]$
		\State $[\indic_A] \gets \textsc{Accessible}([\indic_O], [\indic_{T}], [\indic_{K}], R)$
		\State $[\indic_{P_\o}] \gets [\indic_{P_\i}] - [\indic_A]$
		\EndWhile
		\State \textbf{return} $[\indic_{P_\o}]$
		\EndProcedure
	\end{algorithmic} \label{alg:OC}
\end{algorithm}

\subsubsection{Under-Cut Actions} \label{sec_UC}

As described at the beginning of section \ref{sec_revisit}, a UC action removes all of the material from the region $P_\i - P$, while minimizing the collateral damage, i.e., material removed from $P$ to make all of $P_\i - P$ removable.

Let $P_\o$ represent the output of the UC action whose inputs are the target shape $P$, the current shape (before the action) $P_\i$, the SM tool $T = H \cup K$ with passive part $H$ (e.g., the holder) and active part $K$ (e.g., the cutter), oriented with a fixed rotation $R$. The output pointset must satisfy the following set equation:
\begin{equation}
	P_\o = P_\o - C(P_\o, P, T, K, R). 
	\label{undercut_identity}
\end{equation}
This equation means that the output of the UC action is itself treated as the obstacle pointset $O := P_\o$, against which the collateral region $C(O, P, T, K, R)$ is computed, whose removal from the input state $P_\i$ results in the output state $P_\o$. Once again, this equation can be solved recursively until its fixed point is found.

We begin the iterative process with an initial {\it overestimation} of the obstacle pointset $O_0 := P_\i \cap P$, noting that the UC policy, by definition, will result in an outcome $O = P_\o$ that is contained in $O_0$. However, in order for $O_0 := P_\i \cap P$ to be accessible by the tool, the collateral region $C_0 := C(O_0, P, T, K, R)$ must first be removed; the UC output is thus overestimated as $O_1 := O_0  - C_0$. Similarly, in order for $O_1$ to be accessible, the collateral region $C_1 := C(O_1, P, T, K, R)$ must be removed as well, yielding a slightly better overestimation $O_2 := O_1 - C_1$. Proceeding with this iteration, one ends up with a monotonically decreasing sequence of obstacle pointsets overestimating the UC output, lower-bounded by the true region:
\begin{equation}
	O_0 \supseteq O_1 \supseteq \cdots \supseteq O = \UC(P_\i, P, T, K, R),
\end{equation}
where $P_\o = \UC(P_\i, P, T, K, R)$ represents the converged solution for the UC output. Note that by construction, the fixed point of recursion (i.e., the pointset to which the iteration converges) obeys the identity in \eq{undercut_identity}.

\begin{figure*}
	\centering
	\includegraphics[width=\linewidth]{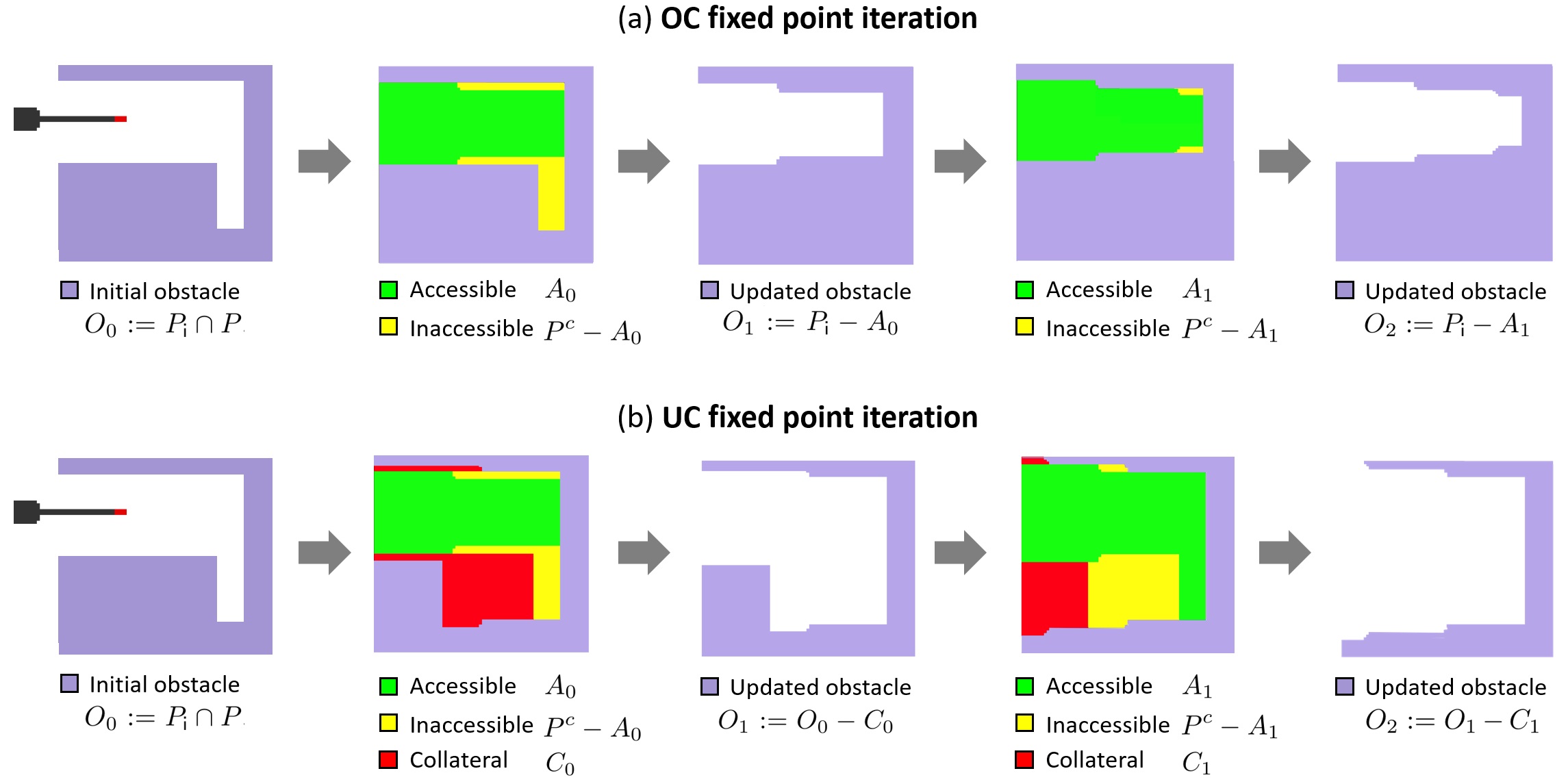}
	\caption{The iterative processes to determine OC and UC regions. Unlike the accessibility analysis in Fig. \ref{fig_access}, one must use an iterative approach to determine what parts of the excess material can be milled since the excess material itself may obstruct the tool from accessing deeper portions of the excess material. The iterative process converges in two steps, in this particular case.}
	\label{fig_iter}
\end{figure*}

Algorithm \ref{alg:UC} describes the recursive method used to compute the UC output. The algorithm uses the subroutine $\textsc{Collateral}([\indic_O], [\indic_{P}],  [\indic_{T}], [\indic_{K}], R)$ from Algorithm \ref{alg:C}. Our experience shows that both iterative procedures converge after a few iterations, as depicted in Fig. \ref{fig_iter}. 

\begin{algorithm}[hbt!]
	\caption{Computing UC output}
	\begin{algorithmic}
		\Procedure{UnderCut}{$[\indic_P]$, $[\indic_{P_\i}]$, $[\indic_{T}]$, $[ \indic_{K}], R$}
		\State $[\indic_O] \gets [\indic_\emptyset]$
		\State $[\indic_{P_\o}] \gets [\indic_P]$
		\While{$[\indic_O] \neq [\indic_{P_\o}]$}
		\State $[\indic_O] \gets [\indic_{P_\o}]$
		\State $[\indic_C] \gets \textsc{Collateral}([\indic_O], [\indic_P], [\indic_{T}], [\indic_{K}], R)$
		\State $[\indic_{P_\o}] \gets [\indic_{P_\i}] - [\indic_C]$
		\EndWhile
		\State \textbf{return} $[\indic_{P_\o}]$
		\EndProcedure
	\end{algorithmic} \label{alg:UC}
\end{algorithm}

\subsection{AM Actions} \label{sec_AM}

With the accessible, inaccessible, and support regions defined, we can precisely formulate the valid AM actions according to our conservative and aggressive policies; namely, under-cut (UC) and over-cut (OC), respectively.

The accessibility constraint is easier to enforce in the case of AM, assuming that the deposition occurs in a bottom-up fashion and the print-head stays on the opposite half-space of what it has deposited at any given time. Hence, it suffices to use the input shape as the obstacle set $O := P_\i$ to obtain the accessible region $A(P_\i, T, K, R)$ from \eq{eq_access} without a need for iteration. As mentioned earlier, for an AM tool $T = H \cup K$, we can ignore collisions of previously deposited layers with the incoming material (abstracted by $K$) in the accessibility analysis, i.e., approximate the accessible region with $A(P_\i, H, \emptyset, R) = (P_\i \oplus (RH))^c = P_\i^c \ominus (RH)$.%
\footnote{A more general formula is $A(P_\i, H, \{\bk\}, R) = (P_\i^c \ominus (RH)) + \bk$ in which the active portion is approximated by a single point where the tool (e.g., nozzle) tip is located. We assume $\bk := (0, 0, 0)$.}

In addition to accessibility, we need to consider support requirements (i.e., overhang avoidance) leading to conservative (UF) and aggressive (OF) policies, explained below. However, there is a nontrivial coupling between support analysis and accessibility. Given a max/min self-supporting pointset $U(P, R)$ or $V(P, R)$ for a build direction defined by $R$, and an accessible pointset $A$, one might think the intersection of the two produces a pointset that is both accessible and max/min self-supporting. However, this is not true, because the intersection may lose the self-supporting property (i.e., have overhangs). To correct for this effect, we need to modify the accessible region as follows, to obtain its maximal subset that is self-supporting:
\begin{equation}
	A^\ast(O, T, K, R) \triangleq U \Big( V(O,R) \cup A(O, T, K, R), R \Big) - V(O,R) . \label{eq_access_mod}
\end{equation}
Note that $A^\ast \subseteq A$, therefore, any point $\bx \in A^\ast$ is not only accessible (i.e., $\bx \in A$) but also supported by a line segment $(\bx + RL) \subset A^\ast$. As a result, this is the maximal subset of $A$ whose intersection with max/min self-supporting pointsets does not spoil their self-supported property. Accordingly, the modified inaccessible region can be computed as:
\begin{equation}
	I^\ast(O, T, K, R) \triangleq O^c - A^\ast(O, T, K, R). \label{eq_inaccess_mod}
\end{equation}

\subsubsection{Under-Fill Action} \label{sec_UF}

As described at the beginning of Section \ref{sec_revisit}, a UF action deposits as much material into the region $P - P_\i$ as possible, while avoiding any material deposition into $P^c - P_\i$.

Let $P_\o$ represent the output of the UF action whose inputs are the target shape $P$, the current shape (before the action) $P_\i$, the AM tool $T = H \cup K$ with passive part $H$ (e.g., the nozzle) and active part $K$ (e.g., the droplet), oriented with a fixed rotation $R$. For the special case in which $P_\i := \emptyset$, the output is given by $P_\o = V(P, R)$. If $P_\i \neq \emptyset$, two major modifications are necessary.

First, a nonempty $P_\i$ imposes an accessibility constraint. We argued earlier that accessibility can be approximated via $A = P^c_\i \ominus (RH)$, which must be modified via \eq{eq_access_mod} to $A^\ast = U(V(P_\i,R) \cup A, R) - V(P_\i,R)$. Hence, the first argument in $P_\o = V(P, R)$ must be replaced with $P \cap A^\ast$ to test the half-line intersection measure in \eq{eq_halfline_min} against the accessible and self-supporting subset $P \cap A^\ast$ of the target shape $P$:
\begin{equation}
	P_\o' \triangleq P_\i \cup U\Big( P \cap A^\ast(P_\i, H, \emptyset, R), R \Big). \label{eq_UF_corr1}
\end{equation}

\begin{figure} [h!]
	\centering
	\centering
	\includegraphics[width=\linewidth]{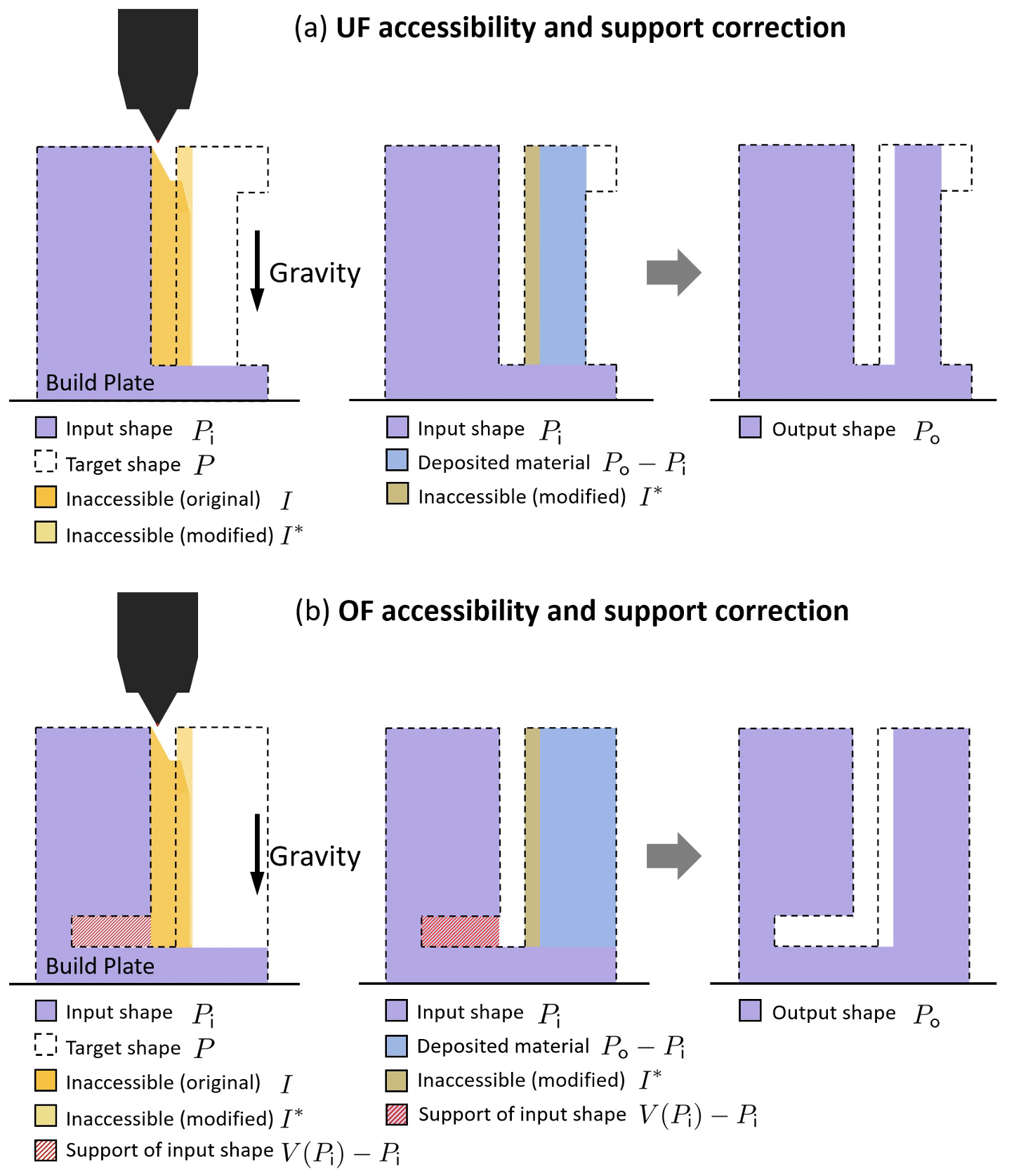}
	\caption{The multi-step process to determine UF and OF regions. Unlike the case of SM actions in Fig. \ref{fig_iter}, the accessibility for AM actions can be computed in on shot based on the input shape. However, the (in)accessible regions must be corrected by underfilling and the input shape must be temporarily modified by overfilling.}
	\label{fig_fill}
\end{figure}

Second, the intersection measure test in \eq{eq_halfline_min} needs to be modified. Assuming the input shape is solidified, there is no need for the entire line segment $\bx + (-RL)$ to fall inside $P \cap A^\ast$. It is sufficient to look only at the portion of the line that is above the input shape $P_\i$, i.e., the overhangs of the input $P_\i$ can be ignored. This effect can be easily achieved by replacing $P \cap A^\ast$ with $(P \cap A^\ast) \cup V(P_\i, R)$ in the argument of the $V$ function, and removing $V(P_\i, R)$ from the result:
\begin{equation}
	P_\o \triangleq P_\i \cup \bigg[ U\Big( \big(P \cap A^\ast(P_\i, H, \emptyset, R)\big) \cup V(P_\i, R), R \Big) - V(P_\i, R) \bigg]. \label{eq_UF_corr2}
\end{equation}
Figure \ref{fig_fill} (a) illustrates how the $U$ and $V$ functions, alongside original and modified accessible and inaccessible regions, are used to compute the UF action's output.

The implicit form of \eq{eq_UF_corr2} is obtained by replacing the Boolean union, intersection, and difference with sign of summation, multiplication, and multiplication with negated value, respectively, of the indicator functions. The discrete form (via 3D voxel arrays) is implemented in Algorithm \ref{alg:UF}.

\begin{algorithm}[hbt!]
	\caption{Computing UF output}
	\begin{algorithmic}
		\Procedure{UnderFill}{$[\indic_P]$, $[\indic_{P_\i}]$, $[\indic_{H}]$, $[ \indic_{K}], R, \bx_\mathrm{B}$}
		\State $[\indic_A] \gets \textsc{Accessible} \big( [\indic_{P_\i}], [\indic_H], [\indic_\emptyset], R \big)$
		\State $[\indic_{A^\ast}] \gets \textsc{MaxSelfSupp}\big( [\indic_A], R, \bx_\mathrm{B} \big)$
		\State $[\indic_{P^\ast}] \gets [\indic_P] \cdot [\indic_{A^\ast}]$
		\State $[\indic_V] \gets \textsc{MinSelfSupp}\big( [\indic_{P_\i}], R, \bx_\mathrm{B} \big)$
		\State $[\indic_E] \gets \sign \big( [\indic_{P^\ast}] + [\indic_V] \big)$
		\State $[\indic_U] \gets \textsc{MaxSelfSupp} \big( [\indic_E], R, \bx_\mathrm{B} \big)$
		\State \textbf{return} $\sign \Big( [\indic_{P_\i}] + \big( [\indic_U] \cdot \neg[\indic_V] \big) \Big)$
		\EndProcedure
	\end{algorithmic} \label{alg:UF}
\end{algorithm}

\subsubsection{Over-Fill Action} \label{eq_OF}

As described at the beginning of section \ref{sec_revisit}, an OF action deposits all of the material into the region $P - P_\i$, while minimizing the sacrificial support, i.e., material deposited into $P^c$ to make all of $P - P_\i$ depositable.

Let $P_\o$ represent the output of the OF action whose inputs are the target shape $P$, the current shape (before the action) $P_\i$, the AM tool $T = H \cup K$ with passive part $H$ (e.g., the nozzle) and active part $K$ (e.g., the droplet), oriented with a fixed rotation $R$. For the special case in which $P_\i := \emptyset$, the output is given by $P_\o = U(P, R)$. If $P_\i \neq \emptyset$, two major modifications are necessary.

First, a nonempty $P_\i$ imposes an accessibility constraint. We argued earlier that accessibility can be approximated via $A = P^c_\i \ominus (RH)$, which must be modified via \eq{eq_access_mod} to $A^\ast = U(V(P_\i,R) \cup A, R) - V(P_\i,R)$. Hence, the first argument in $P_\o = U(P, R)$ must be replaced with $P \cap A^\ast$ to test the half-line intersection measure in \eq{eq_halfline_min} against the accessible and self-supporting subset $P \cap A^\ast$ of the target shape $P$:
\begin{equation}
	P_\o' \triangleq P_\i \cup V\Big( P \cap A^\ast(P_\i, H, \emptyset, R), R \Big). \label{eq_OF_corr1}
\end{equation}

Second, the intersection measure test in \eq{eq_halfline_min} needs to be modified. Assuming the input shape is solidified, there is no need for the entire line segment $\bx + RL$ to fall inside $P \cap A^\ast$. It is sufficient to look only at the portion of the line that is above the input shape $P_\i$, i.e., the overhangs of the input $P_\i$ can be ignored. This effect can be easily achieved by replacing $P \cap A^\ast$ with $(P \cap A^\ast) \cup V(P_\i, R)$ in the argument of the $U$ function, and removing $V(P_\i, R)$ from the result:
\begin{equation}
	P_\o \triangleq P_\i \cup \bigg[ V\Big( \big(P \cap A^\ast(P_\i, H, \emptyset, R)\big) \cup V(P_\i, R), R \Big) - V(P_\i, R) \bigg]. \label{eq_OF_corr2}
\end{equation}
Unlike the case with UF actions, the accessibility constraint in this case often subsumes the intersection test with an upward half-line $\bx + (-RL)$ for query points underneath the overhangs of $P_\i$. More specifically, if $\bx \in A^\ast \subseteq A$ then $\bx + R(T - \bk)$ does not collide with $P_\i$ for all $\bk \in K$. Assuming $(-RL) \subseteq R(T - \bk)$ (i.e., the tool is thicker than a half-line), the half-line does not collide with $P_\i$ either. Therefore, the additional correction in \eq{eq_OF_corr2} is redundant and the simpler formula in \eq{eq_OF_corr1} works (i.e., $P_\o' = P_\o$). 

Figure \ref{fig_fill} (b) illustrates how the $U$ and $V$ functions, alongside original and modified accessible and inaccessible regions, are used to compute the OF action's output.

The implicit form of \eq{eq_OF_corr2} is obtained by replacing the Boolean union, intersection, and difference with sign of summation, multiplication, and multiplication with negated value, respectively, of the indicator functions. The discrete form (via 3D voxel arrays) is implemented in Algorithm \ref{alg:UF}.

\begin{algorithm} [hbt!]
	\caption{Computing OF output}
	\begin{algorithmic}
		\Procedure{OverFill}{$[\indic_P]$, $[\indic_{P_\i}]$, $[\indic_{H}]$, $[ \indic_{K}], R, \bx_\mathrm{B}$}
		\State $[\indic_A] \gets \textsc{Accessible} \big( [\indic_{P_\i}], [\indic_H], [\indic_\emptyset], R \big)$
		\State $[\indic_{A^\ast}] \gets \textsc{MaxSelfSupp}\big( [\indic_A], R, \bx_\mathrm{B} \big)$
		\State $[\indic_{P^\ast}] \gets [\indic_P] \cdot [\indic_{A^\ast}]$
		\State $[\indic_V] \gets \textsc{MinSelfSupp}\big( [\indic_{P_\i}], R, \bx_\mathrm{B} \big)$
		\State $[\indic_E] \gets \sign \big( [\indic_{P^\ast}] + [\indic_V] \big)$
		\State $[\indic_U] \gets \textsc{MinSelfSupp} \big( [\indic_E], R, \bx_\mathrm{B} \big)$
		\State \textbf{return} $\sign \Big( [\indic_{P_\i}] + \big( [\indic_U] \cdot \neg[\indic_V] \big) \Big)$
		\EndProcedure
	\end{algorithmic} \label{alg:OF}
\end{algorithm}

\section{Process Planning} \label{sec_search}

\begin{figure}
	\centering
	\includegraphics[width=\linewidth]{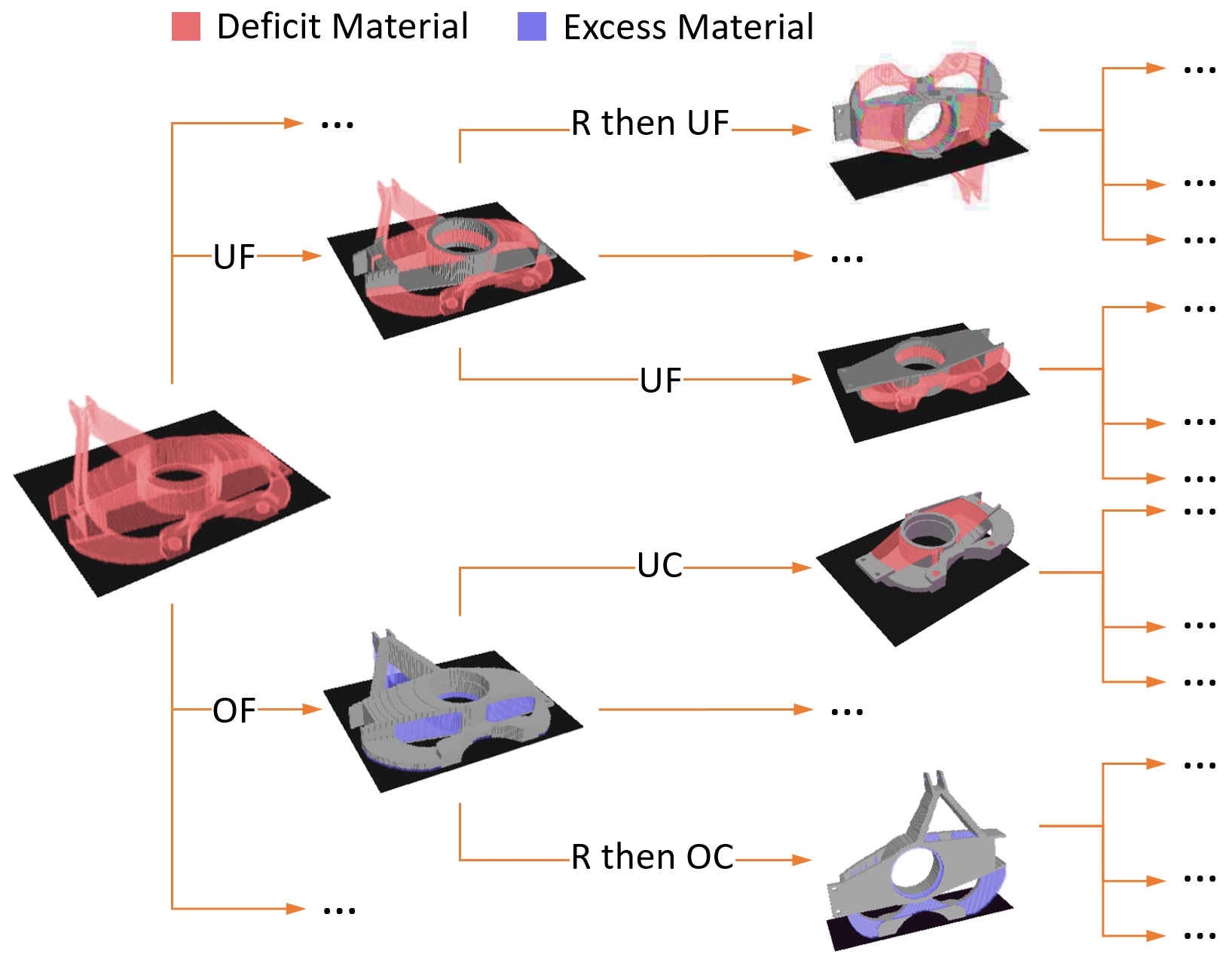}
	\caption{The choice of conservative or aggressive, AM or SM actions leads to a search tree whose paths are valid HM process plans.}
	\label{fig_plan}
\end{figure}

Following \cite{Behandish2018automated}, we define a `valid' HM process plan as a finite sequence of valid AM (UF/OF) and SM (OC/UC) actions $\varphi_1, \varphi_2, \varphi_3, \ldots, \varphi_n$, expressed as a composition of state transitions from an initial part state $P_0$ to a final part state $P_n$ that must be interchangeable (e.g., with respect to a tolerance specification) with the target shape $P$:
\begin{equation}
	P_0 \xrightarrow[]{\varphi_1} P_1 \xrightarrow[]{\varphi_2} P_2 \xrightarrow[]{\varphi_3} \cdots \xrightarrow[]{\varphi_n} P_n \cong P.
\end{equation}
Example of a valid HM process plans are the different branches of the tree depicted in Fig. \ref{fig_plan}.

There are a few difficulties in efficiently exploring the space of potential process plans of arbitrary length $n > 0$. Firstly, there is no easily formulated condition for deciding which manufacturing modality (AM or SM) to use at any given step. In purely AM or SM scenarios, the process is monotonic with each step of the process plan bringing the workpiece closer to the target part. In HM scenarios, monotonicity of the process plan cannot be assumed; for instance, it is possible that removing a feature from the workpiece provides access for a different tool to deposit or remove material elsewhere and that the removed feature may eventually be redeposited at a later stage. Additionally, an infinite number of fixturing/build directions may be possible, so it is necessary to restrict the relative rotations for the search space to be tractable.  

In order to simplify the search space, we assume that the build platform is fixed to be the $xy-$plane and that both AM and SM tools are aligned with the $z-$axis, the part is rotated for each action, and the rotation does not change while the tool and part translate relatively on a $3-$axis motion system. We also restrict the rotations to $90^\circ$ cardinal increments after any given action. Additionally, we assume the tools to be axisymmetric along the $z-$axis, reducing the possible rotation increments at each step to seven in total---namely, the identity, the four $\pm90^\circ$ rotations around the $x-$ or $y-$axes, and the two distinct compositions of these rotations amounting to a single $\pm90^\circ$ rotation around the $x-$ or $y-$axis followed by a $\pm90^\circ$ rotation around the $z-$axis. Finally, we impose the reasonable rule that if at any point the workpiece shape happens to be a subset/superset of the target shape, the next action must be AM/SM respectively. This condition subsumes the special cases where the initial part state is null, hence the first step must be AM, and where the initial state is a raw/bar stock containing the target shape, hence the first step must be SM.

With these restrictions, one can generate a search tree of HM process plans and the resulting as-manufactured part shapes through a sequence of valid AM and SM actions and in-between $\pm90^\circ$ cardinal rotations. Even with a single AM and a single SM tool, the branching factor of the tree can be as large as 28, corresponding to the seven potential cardinal rotations and four manufacturing modalities (OC, UC, UF, OF). Even with these restrictions, the search space quickly becomes intractable, with almost 11,000 potential three-step HM process plans. It should be noted that real-world HM processes may involve choices between multiple different AM and/or SM tools at each step, which increases the branching factor of the search space even further. Therefore, it is crucial to utilize an efficient search algorithm with proper heuristics to manage exponential blowup.

\section{Search Algorithm} \label{se_IDA}

Due to the large search space and the arbitrary geometric complexity of the workpiece at each node, we use an iterative deepening A* (IDA*) search algorithm \cite{Korf1985depth} to determine a cost-optimal HM process plan for a given part. The IDA* search performs a cost-limited depth-first search on the search tree, prioritizing ``promising'' paths (quantified through a cost function) and increasing the maximum allowed cost at each iteration until a solution is found. Compared to a traditional A* search algorithm, The IDA* algorithm (being a depth-first search) is memory efficient since it only has to store nodes on the current stack. Similar to an A* search algorithm, IDA* relies upon a cost function $f = g + h$, where $f$ represents the total cost estimate from the initial state to the goal state, $g$ is the known cost incurred along the current path thus far, and $h$ is the estimated cost from the current node to the goal node, using reasonable heuristics. If the heuristic is admissible, meaning that it is guaranteed to underestimate the actual remaining cost, A* will eventually find the cost-optimal solution \cite{Korf1985depth}.

Our search algorithm terminates once it reaches a goal state, which is defined as arriving at any shape whose excess and deficit regions are small relative to the target shape. We use a simple test for reaching a goal state:%
\footnote{The proper way to do this would be to use interchangeability with respect to form/fit/function, e.g., to check if the excess and deficit regions are restricted to G\&T tolerance zones.}
\begin{equation}
	\frac{\mu^3[P - P_{n}] + \mu^3[P_{n} - P] }{\mu^3[P]} < \delta, \label{eq_interch}
\end{equation}

where $\delta > 0$ is a predetermined small parameter, e.g., our examples use $\delta := 0.01$ unless otherwise stated. This relative error condition provides an adjustable parameter to measure the discrepancy between the as-built and as-designed parts, which allows for termination of the search in case the target part is not manufacturable using any valid combination of the AM/SM capabilities. We generally have $O(10^6)$ active voxels for our test parts, thus allowing for $O(10^4)$ active voxels to be occupied by excess and deficit material. It is also important to note that admitting a nonzero error in the stopping criterion allows for multiple different nodes to be considered terminal.

\subsection{Cost Function} \label{sec_cost}

To construct appropriate cost and heuristic functions, we consider a relatively simple expression for the manufacturing cost of a given part. For a target shape $P$ and an intermediate workpiece state $P_n$ ($n = 0, 1, 2, \ldots$), we can determine how much material (in volumetric terms) must be added to or removed from the as-manufactured part in an ideal scenario to produce the target part. The volume of the deficit material is given by $\mu^3[P - P_n]$ and the volume of the excess material is $\mu^3[P_n - P]$. Note that these are the minimum amount of material that must be moved to get from $P_n$ to $P$, although in reality, additional material may be wasted (i.e., added then removed, or vice versa, in intermediate steps) depending on the geometry, as a result of accessibility and support constraints. Hence, the following formula provides an admissible heuristic:
\begin{equation}
	h(P_{n}, P) \triangleq c^{}_{\mathrm{AM}}\mu^3[P - P_{n}] + c^{}_{\mathrm{SM}}\mu^3[P_{n} - P], \label{eq_heuristic}
\end{equation}
where $c^{}_{\mathrm{AM}}$ and $c^{}_{\mathrm{SM}}$ represent the cost-per-unit-volume of material addition and removal, respectively. One can use a separate coefficient for different AM/SM capabilities, noting the common tradeoff between build/machining time, resolution, roughness, and cost for both AM and SM. However, we use a single pair here for simplicity. Note that, in this case, the ordering imposed by the cost function on the planning space depends on the ratio $\lambda \triangleq c^{}_{\mathrm{SM}}/c^{}_{\mathrm{AM}}$, rather than both parameters at once.

The cost incurred so far, on the other hand, is the sum of costs of all manufacturing actions over preceding workpiece states $P_0, P_1, P_2, \ldots, P_{n-1}$:
\begin{equation}
	g(P_0, P_1, P_2, \ldots , P_n) \triangleq \sum_{i = 0}^{n-1} h(P_{i}, P_{i+1}). \label{eq_cost_sofar}
\end{equation}
in which case, although each action's cost $h(P_{i}, P_{i+1})$ has the same form as the heuristic function in \eq{eq_heuristic}, it is viewed as the actual cost as opposed to an underestimation.

The total cost estimate to arrive from an intermediate workpiece state $P_n$ at a goal state that is close enough to the target state $P$ according to \eq{eq_interch} is thus given by the sum of \eq{eq_heuristic} and \eq{eq_cost_sofar}. 

It is easy to verify that the heuristic is consistent: a workpiece state transition  $P_{i} \xrightarrow[]{\varphi_{i+1}} P_{i+1}$ incurs a cost of at least $h(P_{i}, P_{i+1})$, so the heuristic always provides an underestimate of the cost, and is thus automatically admissible. To speed up convergence of the algorithm, however, we choose a weighted cost function:
\begin{align}
	f_{w}(P_0, P_1, P_2, \dots , P_n, P) &\triangleq g(P_0, P_1, P_2, ..., P_n) \nonumber \\
	&+ (1+w) h(P_n, P), \label{eq_cost}
\end{align}
where $w > 0$ is a parameter determining the weighting of the heuristic.

It is important to note that setting $w > 0$ spoils the admissibility of the heuristic, thus no longer guarantees cost-optimality of the process plan discovered by A*. However, the cost of valid AM/SM actions applied to the workpiece at different directions are often close to each other and the weighting helps distinguish the projected costs of different operations, which is especially critical during the initial stages of the search process.  Additionally, the weight parameter may be used in addition to $\lambda$ to control the search strategy (e.g., OF-then-OC or repeated UF).

\ref{app_cost} provides an analysis of the cost function based on the weighting and relative AM/SM cost factors.

\subsection{Implementation} \label{sec_impl}

We prototyped the IDA* algorithm for HM process planning in \textsf{Python}. In this context, nodes contain information about the workpiece state (including excess and deficit material), manufacturing history including the parent node and all previous operations, and methods to calculate cost incurred so far, heuristic, and total cost-to-goal estimate. The pseudo-code is given in Algorithm \ref{alg:IDA}, in which the subroutines \textsc{Heuristic}, \textsc{TotalCostEst}, and \textsc{Interchangeable} compute the formulae in \eq{eq_heuristic}, \eq{eq_cost}, and \eq{eq_interch}, respectively. The \textsc{Sort} function uses the total cost estimate function to sort the children in ascending order.

\begin{algorithm}[hbt!]
	\caption{IDA* Search for HM Process Planning}
	\begin{algorithmic}
		\Procedure{IDA*}{start, goal} 
			\State found, cost $\gets \textsf{FALSE}, \textsc{Heuristic} \big( \text{start} \big)$
			\While{not found}
				\State found, cost, end $\gets \textsc{Search} \big(\text{start}, \text{goal}, \text{cost} \big)$
			\EndWhile
			\State \textbf{return} cost, end
		\EndProcedure
		\\
		\Procedure{Search}{start, goal, maxCost} 
			\State nodeStack $\gets [\text{start}]$
			\State minCost $\gets \infty$
			\While{nodeStack $\neq \emptyset$}
				\State current $\gets$ nodeStack.pop()
				\If{$\textsf{TotalCostEst} \big( \text{current} \big) \leq \text{maxCost}$}
					\If{$\textsc{Interchangeable}\big($current, goal$\big)$}
					\State $cost \gets \textsf{TotalCostEst} \big( \text{current} \big)$
						\State \textbf{return} \textsf{TRUE}, cost, current
					\Else{}
					\State children $\gets \textsc{GetChildren} \big( \text{current} \big)$
					\State $\textsc{Sort} \big( \text{children}, \textsc{TotalCostEst} \big)$
					\State nodeStack.push(children)
				\EndIf
				\Else{}
					\If{$\textsf{TotalCostEst} \big( \text{current} \big) < \text{minCost}$}
						\State minCost $\gets \textsf{TotalCostEst} \big( \text{current} \big)$
					\EndIf
				\EndIf
			\EndWhile
			\State \textbf{return} \textsf{FALSE}, minCost, null
		\EndProcedure
	\end{algorithmic} \label{alg:IDA}
\end{algorithm}

\section{Results} \label{sec_results}

In this section, we provide examples of HM process plans generated by our algorithm for a support bracket (Section \ref{sec_bracket}), the GE bracket (Section \ref{sec_GE}), a helical screw-thread (Section \ref{sec_threads}), and a helical staircase (Section \ref{sec_stairs}), obtained from \textsf{GrabCAD} and \textsf{Thingiverse}.

\subsection{Support Bracket} \label{sec_bracket}

\begin{figure}
	\centering
	\includegraphics[width=.9\linewidth]{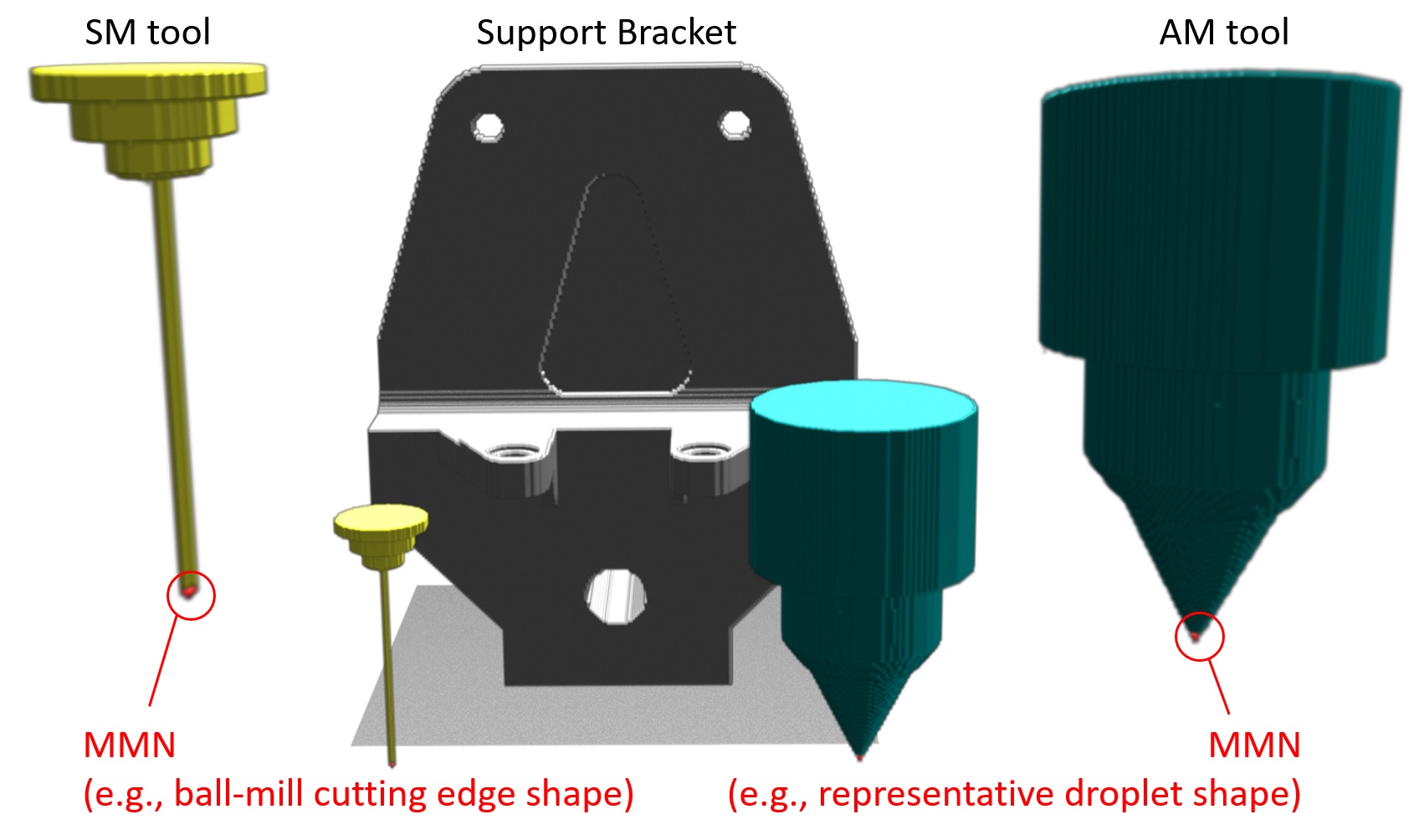}
	\caption{Milling and Additive Manufacturing tools next to support bracket geometry }
	\label{fig: Bracket}
\end{figure}

We consider a support bracket as depicted in Fig. \ref{fig: Bracket} along with an SM tool (ball-end mill) and an AM tool (3D printer nozzle). The target part geometry is voxelized at a resolution of 251$\times$315$\times$243 and has around 4.7 million active voxels. All running times are reported for generating the process plans on a Dell Precision 5820 workstation with and Intel Xeon W-2275 processor, 64 GB DDR4 2933MHz RAM, and an Nvidia  Quadro RTX6000 GPU.

\begin{figure*}
	\centering
	\begin{subfigure}{\linewidth}
		\centering
		\includegraphics[width=\linewidth]{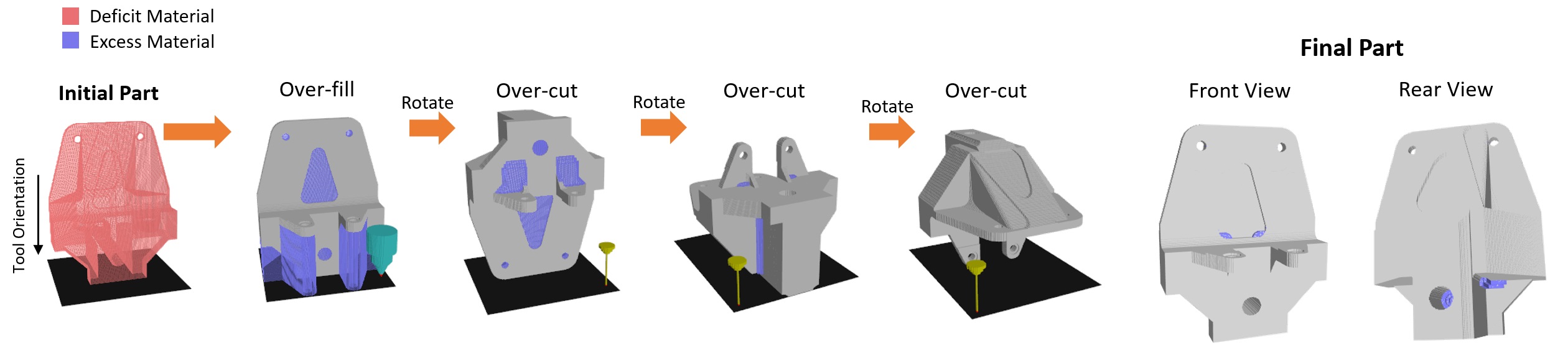}
		\caption{A four-step HM process plan for the support bracket geometry, using $w=1$ and $\lambda := 0.1$.}
		\label{fig_bracket_pp1}
	\end{subfigure}
	\begin{subfigure}{\linewidth}
		\centering
		\includegraphics[width=\linewidth]{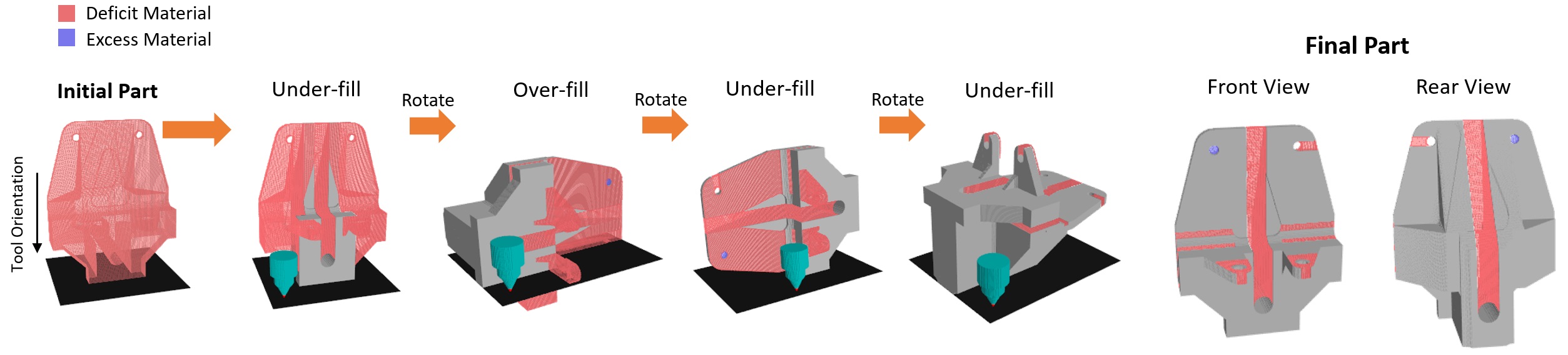}
		\caption{A four-step HM process plan for the support bracket geometry, using $w=1$ and $\lambda := 1$.}
		\label{fig_bracket_pp2}
	\end{subfigure}
	\begin{subfigure}{\linewidth}
		\centering
		\includegraphics[width=\linewidth]{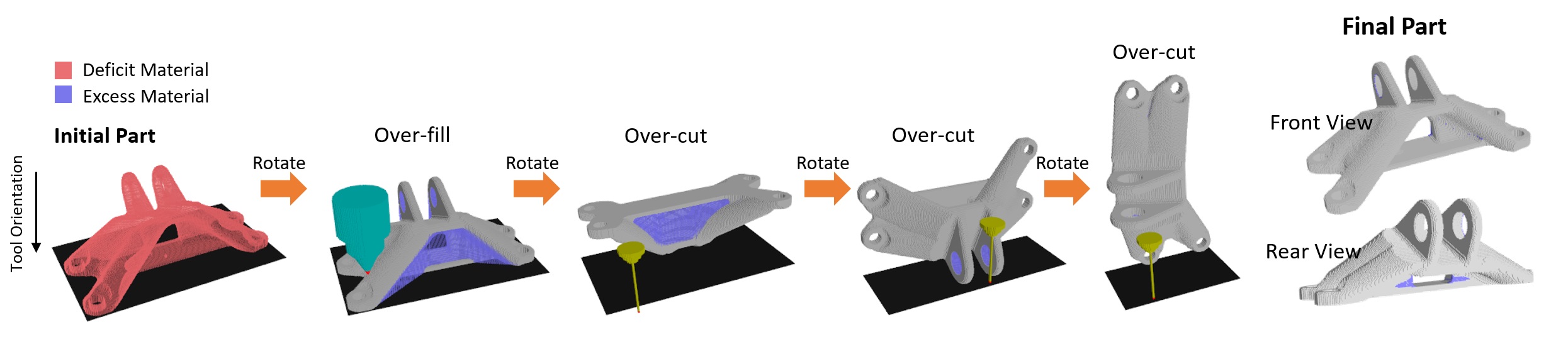}
		\caption{A four-step HM process plan for the optimized GE bracket geometry, using $w=1$ and $\lambda := 0.1$.}
		\label{fig_GE_pp}
	\end{subfigure}
	\begin{subfigure}{\linewidth}
		\centering
		\includegraphics[width=\linewidth]{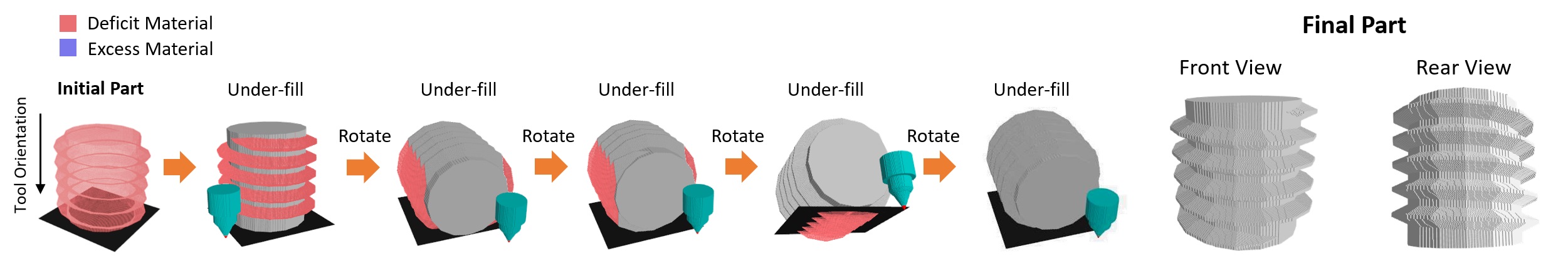}
		\caption{A five-step AM process plan for the helical threaded geometry, using $w=1$ and $\lambda := 0$.}
		\label{fig_threads_pp}
	\end{subfigure}
	\begin{subfigure}{\linewidth}
		\centering
		\includegraphics[width=\linewidth]{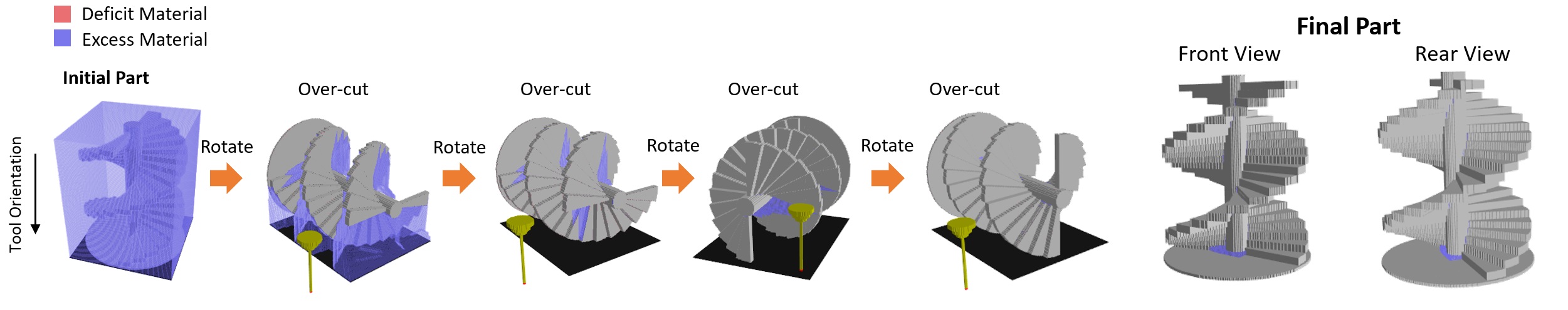}
		\caption{A four-step SM process plan for the helical staircase geometry, using $w=1$ and $\lambda := 0.1$.}
		\label{fig_stairs_pp}
	\end{subfigure}
	\caption{Various HM process plans for the support bracket, topology optimized GE bracket, helical threaded, and helical staircase geometries.}
\end{figure*}

The 4-step process plan identified by our approach is shown in Fig. \ref{fig_bracket_pp1}. For this result, we set the heuristic weight to $w := 1$ and AM-to-SM cost ratio to $\lambda := 0.1$. The relative volumetric error of the as-manufactured shape with respect to the target as-designed shape is $< 0.7 \%$, which amounts to an excess of $\sim3.0 \times 10^{4}$ voxels out of the total $4.7 \times 10^6$ active voxels in the target shape. The actual cost is $\sim 48 \%$ larger than the cost lower-bound, calculated by assuming no (additive or subtractive) material waste. The extra cost is owing to the initial OF action, which deposits excess material that must subsequently be removed. The search process took 1 minute and 41 seconds to complete.

Additionally, we ran our search algorithm with $\lambda := 1$ to better understand how the cost ratio affects the process planning and obtained the 4-step process plan depicted in Fig. \ref{fig_bracket_pp2}. Compared to the previous process plan, this strategy builds the part through a series of UF actions; however, inaccessibility of the AM tool prevents the part from being completely manufacturable through AM alone. No material is wasted in this case, and the cost is $\sim 82 \%$ of the total material cost of the target part. The relative error of the as-manufactured shape is $\sim18 \%$, which translates to a deficit of approximately $8.5 \times 10^{5}$ voxels. This process plan took 4 minutes and 57 seconds to generate.

\subsection{GE Bracket} \label{sec_GE}

Next, we test our algorithm on a topology optimized version of the GE bracket shown in Fig. \ref{fig_GE_pp}. The target part geometry was voxelized at a resolution of 251$\times$93$\times$154 voxels, resulting in around 500K active voxels.

The 4-step process plan identified by our approach is depicted in Fig. \ref{fig_threads_pp}. Again, we set the heuristic weight to $w: = 1$, and AM-to-SM cost ratio to $\lambda := 0.1$. The relative error of the as-manufactured part is $< 0.2 \%$, which amounts to an excess of approximately $\sim 10^3$ voxels. The actual cost is $\sim 45 \%$ larger than the cost lower-bound, owing to the excess material produced by the initial OF action. The search took $21$ seconds to complete.

\subsection{Helical Thread} \label{sec_threads}

The next example is a helical threaded geometry depicted in Fig. \ref{fig_threads_pp}. We voxelized this part at a resolution of 201$\times$141$\times$140, resulting in 2M active voxels, and set $w := 1$ and $\lambda := 0$. For this part, our process planner was able to identify a plan to manufacture the target part in 5 UF steps, with a final error of almost zero. Although the cost ratio was set to zero, the cost of support material for an initial OF operation is large enough that the process planner selects a purely additive strategy. The search took $40$ seconds to complete. 

\subsection{Helical Staircase} \label{sec_stairs}

The last example is a helical staircase geometry depicted in Fig. \ref{fig_stairs_pp}. We voxelized this part at a resolution of 201$\times$163$\times$163, resulting in 460K active voxels, and set $w := 1$, and $\lambda := 0.1$. For this part, we start from a raw stock, and our process planner was able to find a plan to manufacture the target part in 5 OC steps, with a final error of $< 0.9\%$. The search took $30$ seconds to complete.

\section{Conclusion} \label{sec_conclusion}

We presented an approach to define and compute AM/SM actions for arbitrary part and tool geometries and synthesize cost-optimal HM process plans. We proposed conservative and aggressive policies to define AM/SM actions that modify a given workpiece in its intermediate state to generate the closest shape to the given target shape, defined in some sense (e.g., volumetric difference). The conservative (i.e., under-fill AM and over-cut SM) actions do so without producing any excess or deficit regions that require an action of the opposite kind to undo, i.e., get the workpiece closer to the target while assuming the remaining job will require an action of the same kind. The aggressive (i.e., over-fill AM and under-cut SM) actions do so with producing minimal excess or deficit regions, while assuming the remaining job will require an action of the opposite kind to undo their undesired effects. These effects include collateral damage (in the case of under-cut SM), i.e., removing regions that belong to the target shape temporarily to make all regions outside the target shape accessible, and sacrificial material (in the case of over-fill AM), i.e., support structure outside the target shape to carry the weight of all regions inside the target shape. Our approach to define these notions is based on morphological operations on 3D pointsets, and does not make any assumption on the shape, feature-based heuristics, or parametrizations.

The valid AM/SM actions are used to enumerate the space of HM process plans in a tree, which we explore through an IDA* search algorithm to assess manufacturability and generate a cost-effective HM process plan. Even with a fairly crude search heuristic, based on a snapshot of excess and deficit materials, our search algorithm can produce process plans for a variety of complex part geometries. Our cost function provides a crude volumetric measure that may represent material cost and/or machine usage time. In real-world scenarios, however, it may be desirable to select better cost functions which measure other relevant aspects of the manufacturing process including re-fixturing, roughing vs. finishing, etc.

Currently, we do not control error in a spatially localized manner, nor does our stopping criterion consider as-manufactured part performance; however, it is quite natural to desire more refined controls on the form/fit/function or physics-based performance of the as-manufactured part (e.g., based on GD\&T tolerance specifications and simulation-informed performance criteria). Future work may focus on incorporating these considerations into the error measure. 

\section*{Acknowledgments}

The authors are thankful to John T. Maxwell for his insights on hybrid manufacturing and Saigopal Nelaturi for his support of this project under the Design for Manufacturing research program at PARC.

\section*{References}
\bibliography{hybridPolicies}

\begin{thebibliography}{10}
\expandafter\ifx\csname url\endcsname\relax
  \def\url#1{\texttt{#1}}\fi
\expandafter\ifx\csname urlprefix\endcsname\relax\def\urlprefix{URL }\fi
\expandafter\ifx\csname href\endcsname\relax
  \def\href#1#2{#2} \def\path#1{#1}\fi

\bibitem{Rifat2022effect}
M.~Rifat, S.~Basu, E.~De~Meter, G.~Manogharan, Effect of prior surface textures
  on the resulting roughness and residual stress during bead-blasting of
  electron beam melted ti-6al-4v, Crystals 12~(3).
\newblock \href {http://dx.doi.org/10.3390/cryst12030374}
  {\path{doi:10.3390/cryst12030374}}.

\bibitem{Gong2020machining}
Machining Behavior and Material Properties in Additive Manufacturing Ti-6Al-4V
  Parts, Vol.~1 of International Manufacturing Science and Engineering
  Conference.
\newblock \href {http://dx.doi.org/10.1115/MSEC2020-8487}
  {\path{doi:10.1115/MSEC2020-8487}}.

\bibitem{Basinger2018development}
K.~Basinger, C.~Keough, C.~Webster, R.~Wysk, T.~Martin, O.~Harrysson,
  Development of a modular computer-aided process planning (capp) system for
  additive-subtractive hybrid manufacturing of pockets, holes, and flat
  surfaces, The International Journal of Advanced Manufacturing Technology
  96~(5) (2018) 2407--2420.
\newblock \href {http://dx.doi.org/10.1007/s00170-018-1674-x}
  {\path{doi:10.1007/s00170-018-1674-x}}.

\bibitem{Cortina2018latest}
M.~Cortina, J.~Arrizubieta, J.~Ruiz, E.~Ukar, A.~Lamikiz, Latest developments
  in industrial hybrid machine tools that combine additive and subtractive
  operations, Materials 11~(12) (2018) 2583.
\newblock \href {http://dx.doi.org/10.3390/ma11122583}
  {\path{doi:10.3390/ma11122583}}.

\bibitem{Sealy2020hybrid}
M.~Sealy, G.~Madireddy, R.~Williams, P.~Rao, M.~Toursangsaraki, Hybrid
  processes in additive manufacturing, Journal of Manufacturing Science and
  Engineering 140~(6).
\newblock \href {http://dx.doi.org/10.1115/1.4038644}
  {\path{doi:10.1115/1.4038644}}.

\bibitem{Popov2020hybrid}
V.~Popov, A.~Fleisher, Hybrid additive manufacturing of steels and alloys,
  Manufacturing Reviews 7 (2020) 6.
\newblock \href {http://dx.doi.org/10.1051/mfreview/2020005}
  {\path{doi:10.1051/mfreview/2020005}}.

\bibitem{Pragana2021hybrid}
J.~P.~M. Pragana, R.~F.~V. Sampaio, I.~M.~F. Bragança, C.~M.~A. Silva,
  P.~A.~F. Martins, Hybrid metal additive manufacturing: A state-of-the-art
  review, Advances in Industrial and Manufacturing Engineering 2 (2021) 100032.
\newblock \href {http://dx.doi.org/10.1016/j.aime.2021.100032}
  {\path{doi:10.1016/j.aime.2021.100032}}.

\bibitem{Liu2019sequence}
C.~Liu, Y.~Li, S.~Jiang, Z.~Li, K.~Xu, A sequence planning method for five-axis
  hybrid manufacturing of complex structural parts, Proceedings of the
  Institution of Mechanical Engineers, Part B: Journal of Engineering
  Manufacture 234~(3) (2019) 421--430.
\newblock \href {http://dx.doi.org/10.1177/0954405419883052}
  {\path{doi:10.1177/0954405419883052}}.

\bibitem{Chen2019manufacturability}
L.~Chen, T.~Lau, K.~Tang, Manufacturability analysis and process planning for
  additive and subtractive hybrid manufacturing of quasi-rotational parts with
  columnar features, Computer-Aided Design 118 (2019) 102759.
\newblock \href {http://dx.doi.org/10.1016/j.cad.2019.102759}
  {\path{doi:10.1016/j.cad.2019.102759}}.

\bibitem{Zheng2020cost}
Y.~Zheng, J.~Liu, R.~Ahmad, A cost-driven process planning method for hybrid
  additive--subtractive remanufacturing, Journal of Manufacturing Systems 55
  (2020) 248--263.
\newblock \href {http://dx.doi.org/https://doi.org/10.1016/j.jmsy.2020.03.006}
  {\path{doi:https://doi.org/10.1016/j.jmsy.2020.03.006}}.

\bibitem{Behandish2018automated}
M.~Behandish, S.~Nelaturi, J.~Kleer, Automated process planning for hybrid
  manufacturing, Computer-Aided Design 102 (2018) 115--127.
\newblock \href {http://dx.doi.org/10.1016/j.cad.2018.04.022}
  {\path{doi:10.1016/j.cad.2018.04.022}}.

\bibitem{Karunakaran2010low}
K.~P. Karunakaran, S.~Suryakumar, V.~Pushpa, S.~Akula, Low cost integration of
  additive and subtractive processes for hybrid layered manufacturing, Robotics
  and Computer-Integrated Manufacturing 26~(5) (2010) 490--499.
\newblock \href {http://dx.doi.org/10.1016/j.rcim.2010.03.008}
  {\path{doi:10.1016/j.rcim.2010.03.008}}.

\bibitem{Zhang2020development}
W.~Zhang, Development of an additive \& subtractive hybrid manufacturing
  process planning strategy of planar surface for productivity and geometric
  accuracy, International Journal of Advanced Manufacturing Technology 109~(5)
  (2020) 1479--1491.
\newblock \href {http://dx.doi.org/10.1007/s00170-020-05733-9}
  {\path{doi:10.1007/s00170-020-05733-9}}.

\bibitem{Ren2007part}
Part Repairing Using a Hybrid Manufacturing System, International Manufacturing
  Science and Engineering Conference.
\newblock \href {http://dx.doi.org/10.1115/MSEC2007-31003}
  {\path{doi:10.1115/MSEC2007-31003}}.

\bibitem{Wilson2014remanufacturing}
J.~Wilson, C.~Piya, Y.~Shin, F.~Zhao, K.~Ramani, Remanufacturing of turbine
  blades by laser direct deposition with its energy and environmental impact
  analysis, Journal of Cleaner Production 80 (2014) 170--178.
\newblock \href {http://dx.doi.org/10.1016/j.jclepro.2014.05.084}
  {\path{doi:10.1016/j.jclepro.2014.05.084}}.

\bibitem{Yamazaki2016development}
T.~Yamazaki, Development of a hybrid multi-tasking machine tool: Integration of
  additive manufacturing technology with cnc machining, Procedia CIRP 42 (2016)
  81--86, 18th CIRP Conference on Electro Physical and Chemical Machining (ISEM
  XVIII).
\newblock \href {http://dx.doi.org/10.1016/j.procir.2016.02.193}
  {\path{doi:10.1016/j.procir.2016.02.193}}.

\bibitem{Manogharan2015aims}
G.~Manogharan, R.~Wysk, O.~Harrysson, R.~Aman, {AIMS}--a metal additive-hybrid
  manufacturing system: System architecture and attributes, Procedia
  Manufacturing 1 (2015) 273--286.
\newblock \href {http://dx.doi.org/10.1016/j.promfg.2015.09.021}
  {\path{doi:10.1016/j.promfg.2015.09.021}}.

\bibitem{Zhu2012novel}
Z.~Zhu, V.~Dhokia, S.~T. Newman, A novel process planning approach for hybrid
  manufacturing consisting of additive, subtractive and inspection processes,
  in: 2012 IEEE International Conference on Industrial Engineering and
  Engineering Management, Institute of Electrical and Electronics Engineers
  (IEEE), 2012, pp. 1617--1621.
\newblock \href {http://dx.doi.org/10.1109/IEEM.2012.6838020}
  {\path{doi:10.1109/IEEM.2012.6838020}}.

\bibitem{Zhu2013development}
Z.~Zhu, V.~Dhokia, S.~Newman, The development of a novel process planning
  algorithm for an unconstrained hybrid manufacturing process, Journal of
  Manufacturing Processes 15 (2013) 404--413.
\newblock \href {http://dx.doi.org/10.1016/j.jmapro.2013.06.006}
  {\path{doi:10.1016/j.jmapro.2013.06.006}}.

\bibitem{Newman2015process}
S.~Newman, Z.~Zhu, V.~Dhokia, A.~Shokrani, Process planning for additive and
  subtractive manufacturing technologies, CIRP Annals - Manufacturing
  Technology 64~(1) (2015) 467--470.
\newblock \href {http://dx.doi.org/10.1016/j.cirp.2015.04.109}
  {\path{doi:10.1016/j.cirp.2015.04.109}}.

\bibitem{Mirzendehdel2019exploring}
A.~M. Mirzendehdel, M.~Behandish, S.~Nelaturi, Exploring feasible design spaces
  for heterogeneous constraints, Computer-Aided Design 115 (2019) 323--347.
\newblock \href {http://dx.doi.org/10.1016/j.cad.2019.06.005}
  {\path{doi:10.1016/j.cad.2019.06.005}}.

\bibitem{Nelaturi2015automatic}
S.~Nelaturi, G.~Burton, C.~Fritz, T.~Kurtoglu, Automatic spatial planning for
  machining operations, in: Proceedings of the 2015 IEEE International
  Conference on Automation Science and Engineering (CASE'2015), Institute of
  Electrical and Electronics Engineers (IEEE), 2015, pp. 677--682.
\newblock \href {http://dx.doi.org/10.1109/CoASE.2015.7294158}
  {\path{doi:10.1109/CoASE.2015.7294158}}.

\bibitem{Serra1983image}
J.~Serra, Image Analysis and Mathematical Morphology, Academic Press, Inc.,
  1983.

\bibitem{Roerdink2000group}
J.~B. T.~M. Roerdink, Group morphology, Pattern Recognition 33~(6) (2000)
  877--895.
\newblock \href {http://dx.doi.org/10.1016/S0031-3203(99)00152-1}
  {\path{doi:10.1016/S0031-3203(99)00152-1}}.

\bibitem{Lozano-Perez1983spatial}
T.~Lozano-Perez, Spatial planning: A configuration space approach, IEEE
  Transactions on Computers C-32~(2) (1983) 108--120.
\newblock \href {http://dx.doi.org/10.1109/TC.1983.1676196}
  {\path{doi:10.1109/TC.1983.1676196}}.

\bibitem{Shapiro1997maintenance}
V.~Shapiro, Maintenance of geometric representations through space
  decompositions, International Journal of Computational Geometry \&
  Applications 7 (1997) 383--418.
\newblock \href {http://dx.doi.org/10.1142/S0218195997000247}
  {\path{doi:10.1142/S0218195997000247}}.

\bibitem{Lysenko2010group}
M.~Lysenko, S.~Nelaturi, V.~Shapiro, Group morphology with convolution
  algebras, in: Proceedings of the 14th ACM Symposium on Solid and Physical
  Modeling (SPM'2010), Association of Computing Machinery (ACM), 2010, pp.
  11--22.
\newblock \href {http://dx.doi.org/10.1145/1839778.1839781}
  {\path{doi:10.1145/1839778.1839781}}.

\bibitem{Requicha1980representations}
A.~A.~G. Requicha, Representations for rigid solids: Theory, methods, and
  systems, ACM Computing Surveys 12~(4) (1980) 437--464.
\newblock \href {http://dx.doi.org/10.1145/356827.356833}
  {\path{doi:10.1145/356827.356833}}.

\bibitem{Requicha1978mathematical}
A.~A.~G. Requicha, R.~B. Tilove, Mathematical foundations of constructive solid
  geometry: General topology of closed regular sets, Production automation
  project, technical memo., University of Rochester (June 1978).

\bibitem{Guibas1983kinetic}
L.~Guibas, L.~Ramshaw, J.~Stolfi, A kinetic framework for computational
  geometry, in: The 24th Annual Symposium on Foundations of Computer Science
  (SFCS'1983), 1983, pp. 100--111.
\newblock \href {http://dx.doi.org/10.1109/SFCS.1983.1}
  {\path{doi:10.1109/SFCS.1983.1}}.

\bibitem{Mirzendehdel2020topology}
A.~M. Mirzendehdel, M.~Behandish, S.~Nelaturi, Topology optimization with
  accessibility constraint for multi-axis machining, Computer-Aided Design 122
  (2020) 102825.
\newblock \href {http://dx.doi.org/10.1016/j.cad.2020.102825}
  {\path{doi:10.1016/j.cad.2020.102825}}.

\bibitem{Korf1985depth}
R.~E. Korf, Depth-first iterative-deepening: An optimal admissible tree search,
  Artificial Intelligence 27~(1) (1985) 97--109.
\newblock \href {http://dx.doi.org/10.1016/0004-3702(85)90084-0}
  {\path{doi:10.1016/0004-3702(85)90084-0}}.

\end{thebibliography}

\appendix
\section{Cost Function Analysis} \label{app_cost}

It is instructive to analyze the cost function in order to understand the search algorithm's behavior. Considering a workpiece state transition $P_{n-1} \xrightarrow[]{\varphi_{n}} P_{n}$, resulting from either an AM or SM action, we may consider the change in the cost function resulting from this action:
\begin{equation}
	\Delta f = f(P_0, P_1, ..., P_{n-1}, P_n, P) - f(P_0, P_1, ..., P_{n-1}, P),
	\nonumber
\end{equation}
which may be expressed in the following form:
\begin{align}
	\Delta f = &c^{}_{\mathrm{AM}} \Big( \mu^3[S_2] + (1+w) \mu^3[S_3] - w \mu[S_1] \Big) \\
	+ &c^{}_{\mathrm{SM}} \Big( \mu^3[S_3] + (1+w) \mu^3[S_2] - w \mu^3[S_4] \Big),
	\label{eq_delta}
\end{align}
where:
\begin{align}
	S_1& \triangleq (P_{n} - P_{n-1})\cap P,  \nonumber \\
	S_2& \triangleq (P_{n} - P_{n-1})\cap P^c, \nonumber \\
	S_3& \triangleq (P_{n-1} - P_{n})\cap P, \nonumber \\
	S_4& \triangleq (P_{n-1} - P_{n})\cap P^c. \nonumber
\end{align}
$S_1$ represents the decrease in deficit material, $S_2$ represents the increase in excess material, $S_3$ represents the increase in deficit material, and $S_4$ represents the decrease in excess material, due to the action. Note that for AM actions, $S_3= S_4 = \emptyset$, whereas for SM actions $S_1 = S_2 = \emptyset$. Note also that for the conservative AM (i.e., UF) actions, $S_2 = \emptyset$, too (i.e., only $S_1$ is nonempty), while for the conservative SM (i.e., OC) actions, $S_3 = \emptyset$, too (i.e., only $S_4$ is nonempty).

Since the cost function is symmetric with respect to excess and deficit materials, we may consider the case in which $c^{}_{\mathrm{SM}} \ll c^{}_{\mathrm{AM}}$, without loss of generality. In this case, the change in the cost function may be approximated as:
\begin{align}
	\Delta f &\approx c^{}_{\mathrm{AM}} \Big( \mu^3[S_2] + (1+w) \mu^3[S_3] \Big) \\
	&- w c^{}_{\mathrm{AM}} \mu^3[S_1]  - w c^{}_{\mathrm{SM}} \mu^3[S_4].
	\label{eq:approx_delta}
\end{align}
For AM actions (i.e., $S_3 = S_4 = \emptyset$), \eq{eq:approx_delta} reduces to :
\begin{equation}
	\Delta f_{\mathrm{AM}} \approx c^{}_{\mathrm{AM}} \big( \mu^3[S_2] - w \mu^3[S_1] \big),
	\label{eq:approx_delta_add}
\end{equation}
and for subtractive actions (i.e., $A =B = \emptyset$) it reduces to:
\begin{equation}
	\Delta f_{\mathrm{SM}} \approx (1+w) c^{}_{\mathrm{AM}} \mu^3[S_3] - w c^{}_{\mathrm{SM}} \mu^3[S_4].
	\label{eq:approx_delta_sub}
\end{equation}

Notice that if the initial part state is empty (i.e. $P_{0} = \emptyset$), then $\Delta f = 0$ for all SM actions, hence the search algorithm will attempt to minimize $\Delta f$ through one of the AM actions: UF or OF. While UF actions always produce a $\Delta f_{\mathrm{AM}} < 0$, they may be advantageous over OF actions if the cost of depositing support material is less than the weighted reduction in deficit material. Depending upon the part geometry, a larger weighting parameter will prefer an initial OF action, whereas a smaller weighting parameter will prefer an initial UF action. 

In the case that the weighting parameter $w$ is large enough to produce an initial OF action, $\Delta f = 0$ for any subsequent AM action, hence the following action must be SM. The search will be biased towards choosing the second action as OC, since $\Delta f_{\mathrm{SM}} < 0$ is guaranteed for OC actions, but $w c^{}_{\mathrm{SM}} \ll (1+w)c^{}_{\mathrm{AM}}$, hence any increase in deficit material will be penalized heavily. However, if the weighted increase in deficit material is small relative to the weighted reduction in excess material, the search may choose a UC action, which will require an AM action at a later step to repair the incurred material deficit. Repeating this line of reasoning for subsequent actions results in an OF then OC strategy, leading to an HM process plan with alternating AM and SM actions.

In the case that the weighting parameter $w$ is small enough to produce an initial UF action, $\Delta f = 0$ for any subsequent SM action, hence the following action must be AM. By a similar reasoning as the initial action, if the weighting parameter is small, the search will be biased towards repeating UF actions. However, if the weighted increase in excess material is small relative to the weighted reduction in deficit material, the search may choose an OF action, which will require an SM action at a later step to clear the incurred material excess, leading to a monotonic AM process plan. Additionally, in the case that $c^{}_{\mathrm{AM}} \ll c^{}_{\mathrm{SM}}$, one may repeat the above analysis to show that the search will be biased towards monotonic AM process plans in all cases.

Thus two natural HM strategies emerge from choices of the search parameters: a large weight $w$ and a small $\lambda$ ratio will tend to produce an AM-then-SM (OF-then-OC) strategy, while a small weight $w$ and a large $\lambda$ ratio will tend to produce a monotonic AM strategy. For our experiments, we fix $w = 1$ in order to explore how the variation in $\lambda$ affects the HM process planning strategy.


\end{document}